\documentstyle[12pt,aaspp4]{article}

\newcommand{\Msun}{\mbox{ M}_{\odot}}
\newcommand{\gae}{\mathrel{>\kern-1.0em\lower0.9ex\hbox{$\sim$}}}
\newcommand{\lae}{\mathrel{<\kern-1.0em\lower0.9ex\hbox{$\sim$}}}
\newcommand{\kms}{km~s$^{-1}$}
\newcommand{\NV}{\ion{N}{5}\,$\lambda\lambda\,1238.8,\,1242.8$\ }
\newcommand{\CIV}{\ion{C}{4}\,$\lambda\lambda\,1548,\,1550$\ }

\lefthead{Boroson et al.}
\righthead{Her X-1: Empirical Models}

\begin{document}

\title{Hercules X-1: Empirical Models of UV Emission Lines}

\author{Bram Boroson and Timothy Kallman}
\affil{Goddard Space Flight Center, Greenbelt, MD 20771;
bboroson@falafel.gsfc.nasa.gov, tim@xstar.gsfc.nasa.gov}

\and

\author{Saeqa Dil Vrtilek and John Raymond}
\affil{Harvard-Smithsonian Center for Astrophysics;
svrtilek@cfa.harvard.edu, jraymond@cfa.harvard.edu}

\and

\author{Martin Still and Manuel Bautista}
\affil{Goddard Space Flight Center, Greenbelt, MD 20771;
still@chunky.gsfc.nasa.gov, bautista@gsfc.nasa.gov}

\and

\author{Hannah Quaintrell}
\affil{The Open University, Walton Hall, Milton Keynes MK7 6AA, UK}


\begin{abstract}

The UV emission lines of Hercules~X-1, resolved with the HST GHRS and
STIS, can be divided into broad (FWHM$\approx750$~\kms) and narrow
(FWHM$\approx150$~\kms) components. The broad lines can be unambiguously
identified with emission from an accretion disk which rotates prograde
with the orbit.  The narrow lines, previously identified with the
X-ray illuminated atmosphere of the companion star,
are blueshifted
at both $\phi=0.2$ and $\phi=0.8$ and the line flux at
$\phi=0.2$ is $\approx0.2$ of the flux at $\phi=0.8$.
Line ratio diagnostics show that the density
of the narrow line region is $\log n_{\rm e}=13.4\pm0.2$ and $T_{\rm
e}=1.0\pm0.2\times10^{5}$~K. The symmetry of the eclipse ingress suggests
that the line emission on the surface of the disk is left-right symmetric
relative to the orbit.  Model fits to the \ion{O}{5}, \ion{Si}{4}, and
\ion{He}{2} line profiles agree with this result, but fits to the
\ion{N}{5} lines suggest that the receding side of the disk is brighter.  
We note that there are narrow absorption components in the \ion{N}{5}
lines with blueshifts of $\approx500$~\kms.

\end{abstract}

\section{Introduction}

The transfer of matter from the donor star to the compact object in an
X-ray binary has proved to be unexpectedly difficult to understand.  At
first it would appear that the problem is governed simply by classical
hydrodynamics and the boundary conditions are set by the masses and
separation of the two stars. (The latter quantities are well known in the
case of eclipsing pulsars.) The gross structure of accretion disks and
mass transfer has indeed been described
by theory (Shakura \&\ Sunyaev 1973; Lubow \&\ Shu 1975). However,
difficulties are presented by observations of unexpected behavior, by the
realization that X-ray heating is important even in the presence of strong
viscous heating in the disk interior
(Dubus et al. 1999) and that instabilities from the effects of the
radiation on the dynamics may play an important role (Pringle 1996).
Important physical processes in accretion disks, such as the viscous
transport of angular momentum, the generation of magnetic fields, the
nature of disk coronae and winds (Begelman \&\ McKee 1983; Begelman,
McKee, \&\ Shields, 1983), and turbulence within the disk and in the
interaction with the gas stream from the companion (Armitage \&\ Livio
1998), have been characterized or parameterized, but our understanding 
is still in an exploratory stage and has not
been confirmed by observation.

Several X-ray binaries, including Her~X-1, SS~433, LMC~X-4,
GX~354-0, and Cyg~X-2, vary on
time scales of months (Levine et al. 1996, Kong et
al. 1998).  Cataclysmic variables such as TV~Col, V603~Aql, and TT~Ari
also show periods longer than their orbits (Schwarzenberg-Czerny 1992).  
These variations are usually attributed to a precession of the
accretion disk which periodically blocks our line of sight to the
central X-ray source.  Although the analogy between the disk motion and  
solid-body precession is not 
accurate (Kondo, Wolff, \&\ van Flandern 1983), we will use the
conventional term ``precession'' for the disk wobble.  For the
non-eclipsing systems GX~354-0 and Cyg~X-2, the modulation is gradual and
never leads to total obscuration.
This suggests that either the occulting region is on the order of the same
size as the emitting region or that the X-rays are attenuated by an
optically thin absorber (for example, a disk corona).
Jets precess with the X-ray period in SS~433, and
the X-ray pulse profiles vary with the long-term period in Her~X-1.
These phenomena suggest that any precession or warping includes the outer
accretion disk and extends far into the inner disk.

There have been many attempts to explain the long-term period in
Hercules~X-1, the first X-ray source to have a ``superorbital''
period identified.
Most recently, following work of Iping \&\ Petterson (1990), Pringle
(1996)
showed that warped accretion disks are unstable to the re-radiation
pressure that follows from the non-axisymmetric X-ray illumination by the
central source.  This theory has been applied to warped disks in Active
Galactic Nucleii, X-ray binaries, T~Tauri stars, and disks about Young
Stellar Objects (Maloney, Begelman, \&\ Pringle 1996).  According to the
model, the warp should extend to the inner edge of the disk (where the
disk may even flip over, Pringle 1997).  
Radiation-driven warping has the advantage of
predicting a global disturbance in the disk, in which precession occurs at
a single period at various radii within the disk.  The periods predicted
are of the same order of magnitude as observed periods; however, it is
impossible for the theory to explain the variations among individual
sources, as the period depends on imprecisely known parameters such as the
disk viscosity and accretion efficiency.

Axisymmetric disks and their vertical structure have been investigated
theoretically (Shakura \&\ Sunyaev 1973), and the spectra of the X-rays
reprocessed by the surface have been computed (Vrtilek et al. 1990,
Raymond 1993, Ko \&\ Kallman 1994, van Paradijs \&\ McClintock 1995).  
While there is abundant evidence that disks do reprocess X-rays into
optical and UV line and continuum emission, the reprocessing scenarios
involving an axisymmetric disk require that the disk be flared in order
for the outer disk to have an unobstructed view of the central X-ray
source.  However, the standard demonstration that disks should be flared
assumes an isothermal vertical structure for the disk; without this
unrealistic assumption, disks would be convex (Dubus et al. 1999).

Ultraviolet wavelengths are ideal for investigating reprocessing in X-ray
binaries as they contain strong resonance lines from ions that are
sensitive to X-ray illumination.  Hercules~X-1 is a prime target for study
in the UV, as it is fortuitously far from the Galactic plane
($\approx3$~kpc), so that its interstellar reddening is low
($E(B-V)<0.05$).  Observations with IUE (Dupree et al. 1978, Gursky et al.
1980, Howarth \&\ Wilson 1983; Boyle et al. 1986; Vrtilek, Cheng, \&\
Raymond 1996) and the Faint Object Spectrograph (FOS) aboard HST (Anderson
et al. 1994) show that the resonance emission lines of \NV and \CIV are
much stronger than the most prominent optical emission lines,
He\,II$\lambda4686$ and the Bowen blend at 4640\AA\ (Still et al. 1997).  
Boroson et al. (1996) used the Goddard High Resolution Spectrograph (GHRS)
aboard HST to resolve the \ion{N}{5} lines into broad and narrow
components, which they attributed to the accretion disk and the X-ray
illuminated atmosphere of the companion star, respectively. There was no
evidence for double-peaked profiles typical of accretion disks, and the
doublet ratios suggested conditions intermediate between optically thick
and thin limits. Both double-peaked profiles and optically thick
conditions had been expected on theoretical grounds (Raymond 1993, Ko \&\
Kallman 1994).  UV observations of Sco~X-1 (Kallman, Boroson, \&\ Vrtilek
1998) also show profiles that are not double-peaked and doublet ratios
indicating intermediate optical depth.

Considering the uncertainties in current theory and the wealth of
observational details that have yet to find a satisfactory explanation, we
undertake here an {\it empirical} study of the {\it kinematics} of
accretion in Hercules~X-1, based primarily on the UV line profiles
observed recently with the Space Telescope Imaging Spectrograph (STIS)
aboard the Hubble Space Telescope (HST).  These data were part of a
multiwavelength campaign (Vrtilek et al. 1999) that included simultaneous
X-ray observations with the Rossi X-ray Timing Explorer (RXTE), the
Extreme Ultraviolet Explorer (EUVE), and ground-based optical telescopes.
Simultaneous BeppoSAX observations were obtained independently of our
multiwavelength campaign.
Further multiwavelength observations of Her~X-1 
are scheduled for July 1999 and should provide further tests of the models
presented here.

\section{Observations}

The observations took place during the ``short-on'' phase of the 35-day
cycle of Her~X-1 (Jones, Forman, \&\ Liller 1973).  Introducing the
long-term phase $\Psi=0.0$ at X-ray turn-on, the short-on stage occurs
during $\Psi=0.55-0.75$ (Scott \&\ Leahy 1999).  Using the last turn-on
recorded by the RXTE All Sky Monitor and reported by Scott \&\ Leahy (2
cycles before the start of our observation) and applying a long-term
period of 34.853~days.  This period is exactly 20.5 times the binary
period, in agreement with epoch folding of the ASM data and the
observation that 35-day turn-ons occur at $\phi=0.2$ or $\phi=0.7$ with
equal frequency (Scott \&\ Leahy).  With this long-term period and
ephemeris, we find that the
UV observations spanned $\Psi=0.64$ through $\Psi=0.74$.

The UV observations used the Space Telescope Imaging Spectrograph (STIS)
aboard the Hubble Space Telescope (HST).  The STIS instrument design is
described by Woodgate et al. (1998), and the in-orbit performance of the
STIS is described by Kimble et al. (1998).  The STIS Multianode
Microchannel Array (MAMA) detectors were more sensitive to passage of the
satellite through the South Atlantic Anomaly than expected, 
so that we
could only observe Hercules~X-1 during those HST orbits that did not
intersect the SAA.  We thus had to observe in a pattern of 5 HST orbits
followed by 10 HST orbits without observations.  Furthermore, 5 of our HST
orbits produced no data due to a hardware problem.  This paper discusses
the results of the 15 successful HST orbits in July of 1998 (each orbit
provided $\approx45$ minutes of data).  Light curves of the UV flux
observed with the STIS compared with light curves in the X-ray, EUV,
and optical, are shown in Vrtilek et al (1999).

A log of the STIS observations is shown in Table~1.  For observations
extending into mid-eclipse we used the G140L grating, appropriate for
first-order spectroscopy with resolution
$R=\lambda/2\Delta\lambda=960-1440$ ($200-300$~\kms) in the wavelength
range $1150-1720$\AA.  For these observations we used the ACCUM mode to
accumulate spectra in time intervals of 185 seconds.  

For the other 10 HST orbits, we used the E140M grating for high
resolution echelle spectroscopy.  This provides spectral coverages of,
$1150-1710$\AA\ with a resolving power of $R=45,800$ (6~\kms).  
For these exposures, we used the TIME-TAG mode on the STIS, which stamps
each photon detected with a time accurate to $125$~$\mu$sec.  We show the
average spectrum for $\phi=0.685$ through $\phi=0.764$ in
Figure~\ref{fig:plotall},
and identify emission lines.  There are also many interstellar absorption
features in the spectrum which we will discuss in a later paper (Boroson
et al. 1999).

\section{Gaussian Fits\label{sec:gaussian}}

Before interpreting the spectra, we fit the lines with a sum of two
Gaussians, one broad and one narrow.  In fitting two components, our
motivation is to test the interpretation of Boroson et al. (1996) that
there are two emission regions and that the broad component arises in
the disk and the narrow component on HZ~Her.  Although there is no
compelling reason for the lines to have Gaussian shapes, the fits will
allow us to obtain approximate line strengths, doublet ratios, and Doppler
shifts over the course of the orbit.  As the doublet components and
separate line components overlap, a fit to estimate their separate
contributions is needed.

We added together echelle orders where they overlapped, weighting the
fluxes by the inverse square of the errors, and adding weighted errors in
quadrature.  To fit doublets, we assumed each doublet component has the
same Doppler shift relative to its rest wavelength, and allow a doublet
ratio as a free parameter.

For some lines and some orbital phases, the signal to noise ratio in the
lines only warranted a fit with a single Gaussian.  We also fit a flat
continuum; when we did not obtain adequate fits, we allowed a linear
continuum. We determined errors in our fitted parameters using a bootstrap
method (Press et al. 1992).  Orbital phases at each exposure were
determined using the ephemeris of Deeter et al. (1991).

Examples of the fits to the \NV, \ion{O}{5}$\lambda1371$,
\ion{Si}{4}$\lambda1393$, \ion{He}{2}$\lambda1640$, and \CIV lines are
shown in Figures~\ref{fig:gaussn5} through \ref{fig:gausshe2},
respectively. For the \ion{Si}{4}$\lambda1393$ and \CIV lines, we allow
free parameters corresponding to the optical depth and velocity width of
an interstellar absorption line at $-60$~\kms heliocentric velocity. The
best fit parameters and $\chi^2$ values for all lines are shown in
Table~2. In Table~3 we list the doublet ratios of the \ion{N}{5} broad and
narrow components as a function of orbital phase.  In
Figure~\ref{fig:linevels} we show the orbital variation in line velocity,
and in Figure~\ref{fig:fluxn5} we show the orbital variation in line flux
for the \ion{N}{5} line.

We list the lines at $\phi=0.057$ under ``broad lines'' for reasons that
will become clear in \S\ref{sec:basic} and \S\ref{sec:eclipse}. 

\section{Basic Interpretation\label{sec:basic}}

There are four compelling reasons to identify the broad line region with
the accretion disk about the neutron star, as was done by Boroson et al.
(1996).  First, the width of the lines (average FWHM$=750$~\kms) is most
naturally explained as the result of Keplerian broadening in a disk with
radius $r\approx1-2\times10^{11}$~cm, where the velocity at the disk edge
would be $(GM/r)^{1/2} \sin i \approx300-400$~\kms.  Models predict
accretion disk emission
lines should be double-peaked, with the peaks at $\pm$ the projected
velocity at the edge of the disk (Smak 1981).  The FWHM of such a profile
should thus be about twice the velocity at the disk edge, or
$\approx600-800$~\kms.

Second, the behavior of the broad lines
at phases $\phi=\pm0.1$ is consistent with that expected when HZ~Her
partially
occults a Keplerian disk.  The lines at $\phi=0.057$ are
blue-shifted, as expected if the disk rotates prograde with the binary
orbit, so that the approaching side of the disk emerges first from
eclipse.  The magnitude of the Doppler shift, 400~\kms, rules out the
companion star as the source of this emission.  This explanation is
consistent with the line profiles observed
with the GHRS at $\phi=0.88$ and $\phi=0.91$; the approaching side of the
disk is also the first to enter eclipse. 

A third reason to identify the
broad lines with the disk is the variation of line flux with orbital phase
(Figure~\ref{fig:fluxn5}). The broad lines are roughly constant in flux
from $\phi=0.1$
to $0.2$ and from $\phi=0.65$ through $\phi=0.8$.  The STIS low-resolution
spectra from $\phi=0.90-0.98$ show a steadily decreasing flux in the
lines, consistent with the progressive occultation of a spatially extended
region about the neutron star.  

Finally, evidence that is somewhat less compelling is provided by the
Doppler shifts of the broad lines over the course of the orbit
(Figure~\ref{fig:linevels}). As observed previously with the GHRS, the
broad \ion{N}{5} lines are redshifted at $\phi\approx0.75$ at
$\approx200$~\kms (heliocentric), consistent with the neutron star orbital
velocity of $169$~\kms.  At $\phi\lae0.2$, the broad \ion{N}{5} lines have
shifted towards the blue, to $\approx0$~\kms heliocentric velocity.  
Although this is not as much expected from the orbital motion of the
neutron star, we note that the other lines (\ion{Si}{4}, \ion{He}{2},
\ion{C}{4}) are
further blueshifted (Figure~\ref{fig:linevels}).  The
\ion{O}{5}$\lambda1371$ line appears to have either an absorption
component or a separate narrow emission component which renders our line
fits uncertain at phases $\phi<0.211$.  We suspect that scattered emission
(to be discussed in Boroson et al. 1999) may be affecting the fits.  We
also note that
at $\phi=0.211$, there appears to be emission towards the blue that is in
excess of the Gaussian fit to the \ion{N}{5} line.

We identify the narrow line region with the X-ray heated atmosphere of
HZ~Her, as we did in Boroson et al. (1996).  Our principal reason for this
identification (following Boyle et al. 1986) is that the orbital phase
dependence of the line flux resembles the phase dependence of the UV
continuum (Figure~\ref{fig:iueline}), most of which is assumed to arise in
the X-ray heated face of the companion star.  The flux increases from
$\phi=0.1$ to $\phi=0.4$ and decreases from $\phi=0.6$ to $\phi=0.9$ as
the projected area of the X-ray heated atmosphere increases and then
decreases, but between $\phi=0.4$ and $\phi=0.6$ there is a plateau
or secondary minimum as the accretion disk occults the star.  The total
flux should be the total of the flux from the disk and the flux from
HZ~Her, and the
former should be roughly constant over the orbit outside of eclipse.
Difficulties with this picture are highlighted in \S\ref{sec:discuss}.

\section{Eclipse Maps and Broad Line Fits\label{sec:eclipse}}

\subsection{Low-resolution maps of line flux}

The eclipse ingress, observed with the STIS in low-resolution mode, can be
used as a direct test of the left-right asymmetry (relative to the
orbital plane) of the emission in the disk.  This can be a useful
test of the hypothesis that the disk is warped, although if the line
emission is in response to X-rays scattered by a corona, a warped disk can
still produce line emission that is axisymmetric on its surface.

Given masses and separations of the two stars, we can find at a given
orbital phase $\phi$ the distance $x$ from the center of the accretion
disk beyond which the disk is occulted by the surface of the star.
We assume that the eclipsing edge
of HZ~Her is flat relative to our view of the accretion disk.  We have
tested this simplifying assumption by using a model similar to that of
Howarth \&\ Wilson (1983a) to find the actual shape of the eclipsing edge.
We find that the actual eclipse edge has an rms deviation from a
straight line of $\lae5$\%\ of
the radius of the outer edge of the disk.

If the disk is symmetric, then we expect that
\begin{equation}
F_{\rm disk} - F(x) = F(-x)
\end{equation}
That is, the amount of flux occulted when the shadow line passes within
$x$ of the neutron star is the same as the amount remaining when the
shadow line has passed the same distance to the other side of the neutron
star.

This suggests that if we were to plot $(x,F_{\rm disk}-F(x[\phi]))$ and
$(-x[\phi],F(-x[\phi]))$ that they would overlap, given axisymmetric
emission on the disk.  We show the results of this test in
Figure~\ref{fig:symmetry}, for the
continuum flux, for the flux in the \ion{N}{5} lines, and for the flux in
the \ion{C}{4} lines.  The results are consistent with a disk that has
left-right symmetry.  The asymmetry in the \ion{C}{4} lines can be
attributed to the interstellar absorption lines observed with the STIS in
echelle mode.  The absorption component at 1548\AA\ is stronger than the
component at 1550\AA, so that we expect to observe less flux from
\ion{C}{4} when the broad line is blue-shifted, and more of the 1550\AA\
line overlaps the 1548\AA\ interstellar absorption line. The emission
remaining after the disk has started to become eclipsed becomes
progressively more blue-shifted, so that more of the 1550\AA\ emission
line overlaps with the 1548\AA\ interstellar absorption line.

\subsection{Fits to line profiles}

We also fit the line profiles at $\phi=0.057,0.092$, and
$0.132$, phases during the eclipse egress when the narrow line is not
prominent.  An axisymmetric distribution of emission leads to a
double-peaked profile, which does not fit the observation.  We assume
instead that the surface brightness of the disk at
radius $r$ and angle $\theta$ is given by
\begin{equation}
F(r,\theta)=r^\alpha B(\theta)
\end{equation}
where $\alpha$ is a free parameter.  We have set as free parameters 
$B(\theta)$ at 3 equally spaced values of $\theta$ from 0 to $\pi$ (that 
is, in $60^\circ$ intervals; we
force the disk to have front-back symmetry, as the eclipse progression
does not allow us to distinguish between front and back of the disk.)  
We linearly interpolate $B(\theta)$ for values of $\theta$ between
each of the 3 fixed points.
We allow the emission from a point in the disk to be spread into a
Gaussian with velocity width $\Delta v$.  The doublet ratio and $\Delta v$
are both free parameters.  Whether a point on the disk contributes
emission at a given $\phi$ depends on its position relative to the eclipse
line.  We fix $M \sin i$ to be $1.4 \Msun$.  We transform the velocities
of our computed profiles to account for the neutron star's orbit about the
center of mass using
$v^\prime=v-(169$\kms$)\sin 2 \pi \phi$.

We show the results of fits to the \NV doublet in
Figure~\ref{fig:lineprof}.  The best-fit value of $\alpha=-1.5$, the
best-fit disk radius is $1.8\times10^{11}$~cm, the doublet ratio $1.46$,
the turbulent velocity $\Delta v=130$~\kms, and the emissivity versus
angle on the disk is $B(\theta)=(6.6,5.8,1.0)$ for $\theta=(0,\pi/2,\pi)$.
If we exclude from the fit regions near the narrow absorption lines at
$-500$~\kms (see \S\ref{sec:abs}), we still find that the red-shifted side
of the disk is brighter: $B(\theta)=(3.0,1.2,1.0)$.  

The other lines show different best-fit angular brightness profiles,
although the counting statistics are not as good for these lines as for
\NV.  The
\ion{Si}{4}$\lambda1394$ line is fit with similar parameters as the
\ion{N}{5} line, but $B(\theta)=(1.7,0.5,1.0)$.  For \ion{He}{2}, we
find $B(\theta)=(1.9,1.9,1.0)$, and for
\ion{O}{5}$\lambda1371$, we find $B(\theta)=(1.2,0.64,1.0)$.  The best-fit
outer radius for the disk from the \ion{O}{5} profile is $10^{11}$~cm.
It is reasonable to expect that \ion{O}{5}$\lambda1371$ is emitted at
smaller
radii in the disk than the other lines, as it is preferentially emitted at
high
densities.  The low-resolution spectra during eclipse ingress
show that \ion{O}{5} emission disappears at $\phi=0.95$, while the
other emission lines are present.  At this orbital phase, 
only regions further than
$7\times10^{10}$~cm from the neutron star are visible.

We also compare the predictions of the best-fit
model for the \ion{N}{5} lines with the broad lines seen at $\phi=0.88$
and $\phi=0.91$
with the GHRS
(Boroson et al. 1996). To fit the GHRS lines, we fix the parameters
determined by the fits to the STIS spectra, except for the normalization
of the line flux, and allow all velocities to be
multiplied by some factor.  We find the $\chi^2$ deviation from the data 
is minimized when the fluxes are multiplied by 1.5
and the velocities are multiplied by 0.94.  

\section{Density and Temperature Diagnostics}

There are several lines (principally due to O and S) in the UV spectrum
observed with STIS that can be used to examine the density and temperature
of the emitting gas.  To aid our analysis of the line ratios, we use 
the atomic data of the development version of XSTAR, a photoionization
code based on Kallman \&\ McCray (1982).


As suggested by previous observers (e.g. Howarth \&\ Wilson 1983b), the
\ion{Si}{4}$\lambda1403$ line is blended with a complex of \ion{O}{4}
lines.  With the echelle mode of the HST STIS, we can now resolve the
separate lines of this blend. These lines have been used previously in
solar astronomy (Cook et al. 1995) as a density diagnostic.  Nearby
\ion{S}{4} lines at $1423.8$ and $1416.9$\AA\ lines also provide a density
diagnostic (Dufton et al. 1982).

In Figure~\ref{fig:o4lines} we show the average STIS spectra between
$\lambda=1390$ and $\lambda=1430$ for $\phi=0.65$ through $\phi=0.80$.  
This is the phase range in which the narrow lines are most prominent;
because the wavelength separation of the lines is $\sim400$\kms, we are
not able to separate them
at phases when only the broad components are visible.  Narrow lines are
also easier to detect against a noisy background than broad lines.  
The \ion{O}{4} lines at $1338,1343$\AA\ are, surprisingly, about twice as
broad as the narrow \ion{N}{5} components, but the line flux varies over
orbital phase in a manner similar to the narrow lines.  Thus the line
ratios we form should allow us to infer physical
conditions in the narrow line region only.

We estimate the errors in our measurement
of the flux in the lines solely from the counting statistics.
Realistically, errors are also introduced by our choice of continuum, but
these are difficult to characterize.

We can also form a density diagnostic from the ratio of
\ion{O}{5}$\lambda1371$ and \ion{O}{5}$\lambda1218$.  Although
\ion{O}{5}$\lambda1218$ is diminished by the saturated Ly$\alpha$
interstellar absorption line, we correct for this by fitting the
absorption profile to a Lorentzian.  We first adjust the continuum level
so that the saturated flux does not go below 0, to correct for a possible
systematic error in the background subtraction.  We then find an
interstellar column density of $N_{\rm
H}=1.0\pm0.1\times10^{20}$~cm$^{-2}$.  This is within a factor of 2 of
values derived from the X-ray absorption (Vrtilek \&\ Halpern 1985 found
$N_{\rm H}=4.6\times10^{19}$~cm$^{-2}$ absorbing the blackbody flux seen
with Einstein;  Mavromatakis 1993 found $N_{\rm
H}=1.1\pm^{0.4}_{-0.7}\times10^{20}$~cm$^{-2}$ with ROSAT, and Dal Fiume
et al. 1998 found $N_{H}=5.1\pm0.07\times10^{19}$~cm$^{-2}$ with
BeppoSAX.) For standard relations between E(B-V) and $N_{\rm H}$ (Bohlin
1975), we find E(B-V)$=0.018\pm0.002$. From the model of the interstellar
Ly$\alpha$ line, we find we must adjust the observed flux in the
\ion{O}{5}$\lambda1218$ line by a factor of~2.

We obtain the following values for the density diagnostic line ratios:
\ion{S}{4}$\lambda1423/1417=0.9\pm0.3$,
\ion{O}{4}$\lambda 1407/1401=0.6\pm0.2$, \ion{O}{4}$\lambda
1407/1339=0.6\pm0.2$, \ion{O}{4}$\lambda 1401/1343=0.9\pm0.3$,
\ion{O}{4}$\lambda1407/1343=0.5\pm0.2$,
\ion{O}{5}$\lambda1218/1371=0.42\pm0.07$, and \ion{S}{5}$\lambda
1199/1502>6$.  (The \ion{S}{5}$\lambda1199$ line overlaps interstellar
\ion{N}{1} absorption).

In Figure~\ref{fig:manuel} we show the densities and temperatures implied
by these line ratios.  The best agreement for the various line ratios is
found at $\log n_{\rm e}=13.4\pm0.2$, $T_{\rm e}=1.0\pm0.2\times10^{5}$~K.
We note that the \ion{S}{4} ratio is not very sensitive to the
temperature (for a $1\sigma$ change in the ratio, $T_{\rm
e}=4\times10^{5}$~K is consistent with a density of $\log n_{\rm
e}=12.5$.)  However, the \ion{S}{4} ratio can rule out $T_{\rm
e}>6\times10^{5}$~K, and we do not expect temperatures $T_{\rm
e}>2\times10^{5}$~K in any case, as even in coronal ionization balance,
the fraction of \ion{O}{4} ions should be negligable.  The \ion{S}{5}
ratio predicts lower temperatures than the other ratios.  Even lower
temperatures would be predicted if, as we suspect, the flux in the
$\lambda1199$ line is diminished by interstellar \ion{N}{1} absorption.
Recombination from \ion{S}{6} may make a small contribution to the
\ion{S}{5} emission, however, bringing the temperatures into agreement
with those predicted by the \ion{O}{4} and \ion{O}{5} ratios.
Our results for the density are similar to those of Howarth \&\ Wilson
(1983b), who found
$\log n_{\rm e}=13.3$ and $T=2.5\times10^{4}$~K.  The temperature they
found does not appear to be consistent with the \ion{O}{4} ratios,
although it does agree with the \ion{S}{5} ratio.

From the measured flux in the \ion{O}{4}, \ion{O}{5}, and \ion{S}{4},
\ion{S}{5} lines, the measured density, and from the emissivities of the
ions at the measured temperature, we can estimate the emitting volume.  We
assume that all O is in either \ion{O}{4} or \ion{O}{5} stages, and that
all S is in either \ion{S}{4} or \ion{S}{5}.  From the line ratio
\ion{S}{4}$\lambda1424$/\ion{S}{5}$\lambda1502$ we find
$N($\ion{S}{4}$)/N($\ion{S}{5}$)=18\pm4$ (the value in collisional
ionization
balance should be $900$).  From \ion{O}{4}$\lambda
1338$/\ion{O}{5}$\lambda1218$ we find
$N($\ion{O}{4}$)/N($\ion{O}{5}$)=1.7\pm0.5$ (the value for collisional
ionization is 240).  This is further evidence that the gas is in
photoionization equilibrium at $T\lae10^{5}$~K.  Assuming a distance to
Her~X-1 of 5.5~kpc, both the O and S lines imply
$V\approx10^{24}$~cm$^3$.
Such a volume of emitting gas is easy to accomodate
within the system.

Observations with the Hopkins Ultraviolet Telescope
(Boroson et al. 1997) during orbital phases when the narrow lines are
prominent
show that the flux in the \ion{O}{6}$\lambda \lambda1032,1038$ lines 
is approximately equal to that in the \NV lines.  
Using the volume inferred from the O lines, the \ion{O}{6} doublet flux
inferred
from the \NV doublet, and the \ion{O}{6} emissivity, we find that
$\approx8\times10^{-3}$ of the O is in \ion{O}{6}.  This supports our
assumption that all O is in \ion{O}{4} or \ion{O}{5}.

\section{Absorption Lines\label{sec:abs}}

The \ion{N}{5} lines show narrow absorption features at blueshifts of
500\kms, with velocity widths of $b\approx50$\kms
(Figure~\ref{fig:abslines}) and equivalent widths
$W_\lambda\approx0.1$~\AA.  There is also evidence for absorption lines at
$-500$~\kms in \ion{C}{4}$\lambda1548.195$ ($W_\lambda\sim0.2$~\AA, only
seen at $\phi=0.057$).

It is unlikely that these features are artifacts, because they 
persist for several HST orbits and are separated by exactly the doublet
separation in \ion{N}{5}.
The lines are probably not interstellar.
The right-hand panel of Figure~\ref{fig:abslines} shows that as the flux
increases as Her X-1 emerges from eclipse, the lines do not become
significantly stronger, as they would if a constant optical depth were
applied to the changing flux.

This suggests that the absorbing gas may only be in front of the emitting
region which is visible at $\phi=0.057$.  For example, a region above
the front of the disk could absorb emission from the back of the disk.
The problem with the scenario linking the absorption lines to a region of
the disk is that the neutron star line-of-sight velocity, and thus
velocities on the accretion disk, should vary by $\approx100$~\kms from
$\phi=0.057$ to $\phi=0.171$, and yet the absorption feature remains
stationary. 


\section{Discussion\label{sec:discuss}}

One surprising result of the analysis of the eclipse ingress is that the
disk appears to be roughly left-right symmetric in its continuum and
\ion{N}{5} emission.  The analysis that leads to this conclusion, however,
depends on our knowledge of the position of the eclipsing limb of HZ~Her,
that the disk is only occulted by this limb and not by, for example, the
gas stream, on our assumed value for the total flux of the disk, that the
disk emission is constant over the course of the ingress, and that there
is no other variable source of line or continuum emission during the
eclipse ingress.  It would be more satisfying to compare the flux at say
$\phi=0.94$ during one eclipse ingress with the flux at $\phi=0.06$ during
the eclipse egress immediately following.

The disk line profiles are not generally double-peaked.  We can fit the
profiles if we make the empirical assumption that the emission varies with
angle on the disk surface.  In contrast to what we found from the
integrated line flux during eclipse ingress, the line profiles during
egress were fit if the receding side of the disk was brighter than the
approaching edge (Figure~\ref{fig:lineprof}).  Thus the fitting routine
solved the problem of the single-peaked profile by brightening the red
peak at the expense of the blue peak.  The problem of single-peaked
profiles from accretion disks has also been encountered in the study of
AGN.  One possible explantion for red-shifted single-peaked lines is that
gradients in the disk velocities favor the escape of photons from one side
of the disk (Murray \&\ Chiang 1997) into the line of sight.

That our model correctly matches the variation in the line profiles during
both eclipse ingress (observed during the main-on state with the GHRS) and
egress (observed during the short-on state with the STIS) should be
considered strong confirmation of the interpretation that the broad lines
arise in a disk rotating prograde with the orbit and with Keplerian
velocities.  This is the first determination of the rotation direction of
an accretion disk in an X-ray binary.  As X-ray pulsars, including
Her~X-1, are known to change the sign of their spin period derivative
(Nelson et al. 1997), it has been speculated that counter-rotating
accretion disks are possible. That the broad line flux was 50\%\ greater
during the main-on state suggests that the disk is more highly inclined to
the line of sight during the short-on state, as expected, but more
observations are required to separate the 35-day phase variability from
random fluctuations.

If we calculate from this model the eclipse ingress light curve,
the result does not fit the observations.  We
offer two possible explanations for the inconsistency between the eclipse
ingress light curve, which suggests a symmetric disk, and the eclipse
egress line profiles, which do not.
One is that during
the interval of $\approx$~2 days between the ingress and egress, the disk
precessed enough to alter the pattern of illumination and visibility on
its surface.  

Another possibility is that there is another source of emission in
addition to the disk which is complicating the analysis.  For example, the
narrow lines at maximum light are $\approx4$~times brighter than the broad
disk lines. For the \ion{N}{5} lines, $A/n_e C_{21}\sim100$, so that UV
line emission from the star impinging on the disk is much more likely to
be scattered than absorbed. The scattered line profiles would depend on
phase, as the scattering would occur primarily from those points on the
disk with a velocity of $0$ along the line to the narrow line region.  
(The actual velocity of the narrow line is not exactly $0$, but it is less
than the velocities in the disk.) The velocity of these points relative to
the viewer would be $0$ near $\phi=0$ (giving rise to symmetric scattered
emission on the disk) but would cause double-peaked profiles at
$\phi=0.25$ and $\phi=0.75$ (the relative strengths of the peaks would
depend on the geometry). We expect the \ion{N}{5} line to be affected more
than the other lines, as the scattered flux depends on both the strength
of the narrow line flux (\NV is the strongest narrow line) and the optical
depth of the ion in the disk (\ion{N}{5} should have the highest optical
depth, and \ion{He}{2}$\lambda1640$ and \ion{O}{5}$\lambda1371$, which are
not resonance lines, should have the lowest optical depths.)  We note that
the red-shift of the broad \NV lines as determined by the Gaussian fits is
greater at $\phi\approx0.2$ than the redshift determined for the other
lines.
The flux of the scattered lines would depend on the projected area of the
disk as seen by the star and on the local turbulence in the disk. We will
discuss models of these lines in a future paper (Boroson et al. 1999).

We have suggested (Boroson et al. 1996) that the narrow line emission is
due to the X-ray
illuminated atmosphere of HZ~Her, as this can explain the gross behavior
of the line flux as a function of $\phi$.  In addition, the narrow line
flux was approximately equal during the main-on observations observed with
the GHRS (Boroson et al. 1996) and during the short-on observations
reported here. The disk-formed broad lines were brighter during the
main-on state.

There are two aspects of the current observations which cast doubt
on whether the region that causes the narrow line emission is really
the X-ray illuminated stellar atmosphere.
First, the velocities of the narrow lines do not match the
velocities expected from the visible, heated portion of HZ~Her.  This was
apparent in the GHRS observations presented by Boroson et al. (1996).  For
the current STIS observations, we find that the narrow lines are
blueshifted both at $\phi\approx0.75$ and at $\phi\approx0.2$.  

One possibility that we have considered to explain the narrow line
velocities is that there is a gas flow in the atmosphere of HZ~Her
tangential to the Roche surface.  It has been shown that convective
currents resulting from the X-ray heating of HZ~Her should have velocities
of only a few \kms (Dahab 1974), but ablation of a stellar atmosphere by
radiation pressure has been shown to be effective by Voit (1990). Ablation
of HZ~Her could cause blueshifted lines at $\phi\approx0.75$ and
$\phi\approx0.2$, but would be difficult to reconcile with the Doppler
maps of optical lines (Quaintrell 1998).  The optical \ion{N}{5} line
arises in a region that moves with a velocity consistent with the heated
face of HZ~Her.  This line is formed by recombination from \ion{N}{6} and
should thus arise in regions {\it hotter} than the \ion{N}{5}$\lambda1240$
region.  Yet the optical absorption lines, which are formed in gas {\it
cooler} than the \ion{N}{5}$\lambda1240$ region, also move with the
expected velocity of HZ~Her (Still et al. 1997, Quaintrell 1998).  It does
not seem likely that the region emitting \ion{N}{5}$\lambda1240$ could
move along the stellar surface at velocities $\sim100$~\kms, while
sandwiched between stationary gas layers.  The model does make the simple
prediction, however, that the narrow lines should be red-shifted at
$\phi=0.5$.

The second problem with identifying the narrow line region with a region
above HZ~Her is that the narrow line flux at $\phi=0.2$ is
$\approx0.2$ of the narrow line flux at $\phi=0.8$.  The IUE observations
(Figure~\ref{fig:iueline}) show that the {\it total} broad and narrow flux
during our observations did not have a phase variation that was unusual
compared with the historical behavior of the source.  Thus it is possible
that the narrow line emission at $\phi=0.2$ is {\it always} $\approx0.2$
of the emission at $\phi=0.8$.

It is possible for the continuum and narrow line to have different phase
dependances, even though they both arise on HZ~Her.  For example, grazing
incidence could enhance line emission but not continuum emission (Boyle et
al. 1986), so that the lines would be brighter at the limb.  The X-ray
shadow of the disk on the star could then cause an orbital asymmetry that
was different for the lines and continuum.  We continue to suggest
provisionally, in the absence of a better candidate region for the narrow
lines, that they arise on HZ~Her.  To be consistent with the data,
however, this identification needs to invoke some unknown details of the
radiative transfer or the dynamics in the atmosphere to account for the
line velocities and the variation of line flux with $\phi$.

Her~X-1 is a complex system, and difficult to unravel in detail.  However,
we have shown that with phase-resolved spectroscopy, we can separate two
emission regions, and can examine
the rotation velocities in the accretion disk.
As more far UV data on this system is obtained, from our multiwavelength
campaign using the HST STIS, and from the Lyman FUSE (Far-Ultraviolet
Spectroscopic Explorer), we can test more
detailed models that include the differences between narrow line and
continuum emission, occultation and emission by the gas stream, and disk
warp shapes suggested by competing theories.

\acknowledgements

Based on observations with the NASA/ESA {\it Hubble Space Telescope},
obtained at the Space Telescope Science Institute, which is operated by
the Association of Universities for Research in Astronomy, Inc., under
NASA contract GO-05874.01-94A.  BB and SDV supported in part by NASA
(NAG5-2532, NAGW-2685), and NSF (DGE-9350074).  BB acknowledges an NRC
postdoctoral associateship.  HQ is employed on PPARC grant L64621.

\clearpage

\figcaption{The UV spectrum of Her~X-1 observed with the HST STIS in
echelle mode, 
averaged over $\phi=0.685-0.764$.}

\figcaption{Gaussian fits to the \ion{N}{5}$\lambda\lambda1238.8,1242.8$ 
doublet as observed with the HST STIS in echelle mode during July, 1998.
The broad and
narrow line fits are shown separately.}

\figcaption{Gaussian fits to the \ion{O}{5}$\lambda1371$ line as
observed with the HST STIS in July, 1998.  We show only the total of the
broad and narrow line fits.}

\figcaption{Gaussian fits to the \ion{Si}{4}$\lambda1393$ line as observed
with the HST STIS in July, 1998.}

\figcaption{Gaussian fits to the
\ion{C}{4}$\lambda\lambda 1548.195,1550.77$ doublet as
observed with the HST STIS in July, 1998.}

\figcaption{Gaussian fits to the \ion{He}{2}$\lambda1640.47$ line as
observed with the HST STIS in July, 1998.}

\figcaption{Velocities of broad (`*' symbols) and narrow ($+$ signs)
lines from Gaussian fits, versus orbital
phase. The
expected velocities
of the neutron star, the center of mass of HZ~Her, and the L1 Lagrangian
point are indicated.}

\figcaption{Orbital phase variations in the fluxes of broad and narrow
\ion{N}{5} line components from STIS observations.}

\figcaption{Flux of the \ion{N}{5}$\lambda1240$ line, the
\ion{C}{4} line, and the continuum (1260\AA\ to 1630\AA) as
observed with IUE, versus
orbital phase.  The diamonds show the fluxes observed with the HST STIS.}

\figcaption{A test of the left-right symmetry of the disk.  The x axis
shows the expected position of the eclipse line of HZ~Her on the disk,
relative to the neutron star.  The `*' points mark disk fluxes observed,
while the `+' points mark the amount of disk flux occulted, reflected
about $x=0$.  We show flux in (a) the \ion{N}{5} lines, (b) the \ion{C}{4}
lines, (c) the continuum (1260\AA\ to 1630\AA).}

\figcaption{Model fits to the \ion{N}{5} emission lines.  Upper panels:
observed and best-fit (bold) fluxes for the July 1998 observations.
Lower panels: observed and best-fit fluxes for the GHRS observations
of August 1994 (Boroson et al. 1996).  Lower right panel: the best-fit
empirical distribution
of flux with angle on the disk.}   

\figcaption{Emission lines in the 1400\AA\ region which can serve as
density diagnostics.  The vertical dashed lines indicate the rest
wavelength of the lines.  The spectrum has been red-shifted to account
for a blue-shift of $\approx50$~\kms in all of the lines.}

\figcaption{The density versus temperature in the narrow line emission
region implied by different line ratios}

\figcaption{Narrow absorption lines blueshifted at 500\kms.  The
vertical dashed 
lines mark $-500$\kms heliocentric velocity in each of the doublet
components.  In panels (a) through (e) respectively, we show the spectrum 
at $\phi=0.057,0.092,0.132,0.171,0.211$.  In the right-hand panels, we
show the differences of the spectra from the spectrum at $\phi=0.057$.}

\clearpage

\begin{deluxetable}{ccccc}
\tablecaption{The STIS observation log}
\tablehead{
\colhead{Root name} & \colhead{STIS mode\tablenotemark{a}} &
\colhead{Start (MJD)} &
\colhead{Exposure (s)} & \colhead{Orbital
Phase\tablenotemark{b}}}
\startdata
\tableline
O4V401010 & L & 51004.563157 & 565 & 0.906\\  
O4V401020 & L & 51004.571792 & 1517 & 0.917\\
O4V401030 & L & 51004.623308 & 1495 & 0.947\\
O4V401040 & L & 51004.646178 & 1702 & 0.962\\ 
O4V401050 & L & 51004.702278 & 444 & 0.987\\
O4V403010 & E & 51006.513968 & 2227 & 0.057\\
O4V403020 & E & 51006.571780 & 2636 & 0.092\\
O4V403030 & E & 51006.638968 & 2636 & 0.132\\
O4V403040 & E & 51006.706155 & 2636 & 0.171\\
O4V403050 & E & 51006.773342 & 2620 & 0.211\\
O4V404010 & E & 51007.522127 & 2227 & 0.650\\
O4V404020 & E & 51007.579581 & 2636 & 0.685\\
O4V404030 & E & 51007.646768 & 2636 & 0.725\\
O4V404040 & E & 51007.713968 & 2636 & 0.764\\
O4V404050 & E & 51007.781155 & 2620 & 0.804\\
\enddata
\tablenotetext{a}{`L' denotes low-resolution observations, and `E' denotes
echelle observations}
\tablenotetext{b}{The orbital phase of the mid-exposure time, using the
ephemeris of Deeter et al. (1991)}
\end{deluxetable}

\clearpage

{\scriptsize
\begin{deluxetable}{ccccccccc}
\tablecaption{Gaussian fits to the UV lines}
\tablehead{\colhead{$\phi$} & \colhead{$V_{\rm n}$\tablenotemark{a}} &
\colhead{$\Delta V_{\rm n}$\tablenotemark{b}} &
\colhead{$F_{\rm n}$\tablenotemark{c}} & \colhead{$V_{\rm broad}$} &
\colhead{$\Delta
V_{\rm b}$} & \colhead{$F_{\rm b}$} 
 & \colhead{$\chi^2_\nu$, $\nu$}}

\startdata

Line: & & & & \ion{N}{5} \\

0.057 & & & & $ -391\pm 371$ & $ 160\pm 279$ & $ 0.40\pm 0.24$ & 2.45, 238\\
0.092 & & & & $ -38\pm 24$ & $ 832\pm 66$ & $ 4.12\pm 0.23$ & 5.95, 238\\
0.132 & & & & $ -15\pm 6$ & $ 854\pm 12$ & $ 4.77\pm 0.06$ & 7.40, 238\\
0.171 & $ -168\pm 188$ & $ 66\pm 18$ & $ 0.25\pm 0.06$ & $ 7\pm 28$ & $ 822\pm 18$ & $ 4.66\pm 0.10$ & 7.32, 234\\
0.211 & $ -126\pm 139$ & $ 65\pm 36$ & $ 0.44\pm 0.12$ & $ 26\pm 17$ & $ 831\pm 36$ & $ 4.99\pm 0.17$ & 7.35, 234\\
0.650 & $ -51\pm 2$ & $ 182\pm 84$ & $ 6.77\pm 0.31$ & $ 207\pm 41$ & $ 569\pm 84$ & $ 4.88\pm 0.51$ & 17.67, 234\\
0.685 & $ -67\pm 1$ & $ 154\pm 127$ & $ 5.43\pm 0.17$ & $ 211\pm 29$ & $ 654\pm 127$ & $ 5.96\pm 0.45$ & 18.82, 234\\
0.725 & $ -73\pm 36$ & $ 140\pm 82$ & $ 4.04\pm 0.17$ & $ 228\pm
32$ & $ 656\pm 82$ & $ 5.85\pm 0.62$ & 14.36, 234\\
0.764 & $ -94\pm 1$ & $ 151\pm 75$ & $ 3.75\pm 0.12$ & $ 224\pm 32$ & $ 702\pm 75$ & $ 5.01\pm 0.35$ & 10.1, 234\\
0.804 & $ -104\pm 112$ & $ 154\pm 127$ & $ 2.66\pm 0.20$ & $ 217\pm 66$ & $ 866\pm 127$ & $ 5.52\pm 0.51$ & 12.21, 234\\

\tableline

Line: & & & & \ion{Si}{4} \\

0.057 & & & & $ -453\pm 16$ & $ 298\pm 37$ & $ 0.16\pm 0.01$ & 0.42, 141\\
0.092 & & & & $ -226\pm 69$ & $ 940\pm 267$ & $ 0.72\pm 0.47$ & 2.7, 141\\
0.132 & & & & $ -202\pm 50$ & $ 889\pm 181$ & $ 0.86\pm 0.38$ & 3.8, 141\\
0.171 & $ -87\pm 8$ & $ 23\pm 42$ & $ 0.02\pm 0.01$ & $ -151\pm 13$ & $ 893\pm 42$ & $ 0.88\pm 0.05$ & 4.17, 137\\
0.211 & $ -129\pm 4$ & $ 68\pm 285$ & $ 0.04\pm 0.01$ & $ -161\pm 17$ & $
937\pm 285$ & $ 0.63\pm 0.21$ & 5.20, 137\\
0.650 & $ -49\pm 3$ & $ 180\pm 6$ & $ 0.71\pm 0.02$ & $ 124\pm 14$ & $ 415\pm 6$ & $ 0.36\pm 0.03$ & 12.72, 137\\
0.685 & $ -78\pm 2$ & $ 145\pm 30$ & $ 0.59\pm 0.01$ & $ 266\pm 13$ & $ 446\pm 30$ & $ 0.33\pm 0.02$ & 6.47, 137\\
0.725 & $ -111\pm 6$ & $ 152\pm 29$ & $ 0.24\pm 0.05$ & $ 131\pm 43$ & $ 541\pm 29$ & $ 0.31\pm 0.04$ & 8.44, 137\\
0.764 & $ -127\pm 2$ & $ 72\pm 24$ & $ 0.16\pm 0.01$ & $ 101\pm 16$ & $ 618\pm 24$ & $ 0.46\pm 0.02$ & 6.1, 137\\
0.804 & $ -113\pm 7$ & $ 286\pm 14$ & $ 0.44\pm 0.02$ & $ 310\pm 23$ & $ 517\pm 14$ & $ 0.33\pm 0.03$ & 4.06, 137\\
\tableline
\tablebreak

Line: & & & & \ion{O}{5} \\

0.057 & & & & $ -429\pm 119$ & $ 628\pm 488$ & $ 0.08\pm 0.10$ & 0.38, 218\\
0.092 & & & & $ 15\pm 145$ & $ 580\pm 510$ & $ 0.30\pm 0.52$ & 2.64, 218\\
0.132 & $ 213\pm 6$ & $ 214\pm 48$ & $ 0.17\pm 0.01$ & $ -667\pm 22$ & $
773\pm 48$ & $ 0.35\pm 0.02$ & 3.81, 214\\
0.171 & $ 164\pm 8$ & $ 284\pm 104$ & $ 0.23\pm 0.01$ & $ -631\pm 46$ & $
393\pm 104$ & $ 0.13\pm 0.02$ & 3.56, 214\\
0.211 & $ -174\pm 24$ & $ 6\pm 0$ & $ 0.0\pm 0.01$ & $ -52\pm 4$ & $
130\pm 0$ & $ 0.14\pm 0.01$ & 5.33, 214\\
0.650 & $ -80\pm 4$ & $ 263\pm 10$ & $ 0.94\pm 0.05$ & $ 335\pm 36$ & $
570\pm 10$ & $ 0.43\pm0.02$ & 7.47, 214\\
0.685 & $ -81\pm 1$ & $ 216\pm 3$ & $ 1.01\pm 0.01$ & $ 351\pm 10$ & 
$ 447\pm 3$ & $ 0.43\pm 0.02$ & 8.27, 214\\
0.725 & $ -77\pm 1$ & $ 168\pm 6$ & $ 0.63\pm 0.01$ & $ 343\pm 7$ & 
$ 383\pm 6$ & $ 0.48\pm 0.02$ & 7.62, 214\\
0.764 & $ -79\pm 2$ & $ 210\pm 21$ & $ 0.51\pm 0.05$ & $ 380\pm 54$ &
 $ 433\pm 21$ & $ 0.30\pm 0.08$ & 5.68, 214\\
0.804 & $ -122\pm 4$ & $ 369\pm 9$ & $ 0.60\pm 0.01$ & $ 495\pm 11$ & $
427\pm 9$ & $ 0.31\pm 0.02$ & 4.50, 214\\

\tableline

Line: & & & & \ion{C}{4} \\
0.057 & & & & $ -400\pm 11$ & $ 322\pm 38$ & $ 0.74\pm 0.09$ & 0.68, 234\\
0.092 & & & & $ -120\pm 19$ & $ 789\pm 33$ & $ 2.65\pm 0.54$ & 1.46, 234\\
0.132 & & & & $ -146\pm 25$ & $ 742\pm 38$ & $ 3.34\pm 0.58$ & 2.71, 234\\
0.171 & $ -160\pm 26$ & $ 54\pm 86$ & $ 0.25\pm 0.15$ & $ -115\pm 30$ & $ 929\pm 86$ & $ 3.98\pm 0.62$ & 2.90, 230\\
0.211 & $ -129\pm 13$ & $ 66\pm 69$ & $ 0.44\pm 0.22$ & $ -186\pm 13$ & $ 915\pm 69$ & $ 5.7\pm 1.1$ & 3.28, 230\\
0.650 & $ -56\pm 2$ & $ 222\pm 88$ & $ 3.93\pm 0.47$ & $ 194\pm 36$ & $ 536\pm 88$ & $ 2.30\pm 0.45$ & 5.38, 230\\
0.685 & $ -73\pm 2$ & $ 190\pm 84$ & $ 2.68\pm 0.44$ & $ 182\pm 29$ & $ 602\pm 84$ & $ 4.65\pm 1.09$ & 4.33, 230\\
0.725 & $ -86\pm 4$ & $ 129\pm 63$ & $ 1.29\pm 0.22$ & $ 251\pm 22$ & $ 655\pm 63$ & $ 3.99\pm 0.37$ & 4.64, 230\\
0.764 & $ -116\pm 8$ & $ 93\pm 57$ & $ 0.94\pm 0.22$ & $ 121\pm 23$ & $ 611\pm 57$ & $ 7.41\pm 1.69$ & 4.83, 230\\
0.804 & $ -131\pm 7$ & $ 108\pm 38$ & $ 0.81\pm 0.21$ & $ 93\pm 27$ & $ 589\pm 38$ & $ 3.26\pm 0.47$ & 3.54, 230\\

\tableline
\tablebreak

Line: & & & & \ion{He}{2} \\

0.057 & & & & $ -282\pm 44$ & $ 412\pm 140$ & $ 0.26\pm 0.08$ & 0.26, 163\\
0.092 & & & & $ -141\pm 27$ & $ 699\pm 120$ & $ 0.83\pm 0.17$ & 1.1, 163\\
0.132 & & & & $ -78\pm 29$ & $ 605\pm 78$ & $ 0.76\pm 0.13$ &
1.43000, 163\\
0.171 &  &  &   & $ -112\pm 16$ & $ 640\pm 33$ & $
0.95\pm 0.06$ & 1.30, 159\\
0.211 & $ -58\pm 7$ & $ 143\pm 63$ & $ 0.24\pm 0.04$ & $ -110\pm 26$ & $ 642\pm 63$ & $ 0.86\pm 0.07$ & 2.01, 159\\
0.650 & $ -67\pm 9$ & $ 220\pm 27$ & $ 1.45\pm 0.18$ & $ 203\pm 49$ & $ 674\pm 27$ & $ 0.91\pm 0.15$ & 3.75, 159\\
0.685 & $ -82\pm 1$ & $ 185\pm 5$ & $ 1.11\pm 0.02$ & $ 109\pm 12$ & $ 620\pm 5$ & $ 1.22\pm 0.09$ & 3.60, 159\\
0.725 & $ -94\pm 4$ & $ 162\pm 13$ & $ 0.75\pm 0.05$ & $ 108\pm 17$ & $ 682\pm 13$ & $ 1.28\pm 0.07$ & 2.89, 159\\
0.764 & $ -89\pm 3$ & $ 165\pm 13$ & $ 0.65\pm 0.05$ & $ 6\pm 26$ & $ 955\pm 13$ & $ 1.75\pm 0.15$ & 2.82, 159\\
0.804 & $ -106\pm 5$ & $ 161\pm 22$ & $ 0.47\pm 0.07$ & $ 35\pm 21$ & $ 705\pm 22$ & $ 1.09\pm 0.09$ & 2.26, 159\\
\tableline

\enddata
\tablenotetext{a}{Heliocentric velocity of the narrow line component in
\kms}
\tablenotetext{b}{Full Width Half Maximum of fit to the narrow line
component in \kms}
\tablenotetext{c}{Flux in the narrow line, in
10$^{-13}$~erg~s$^{-1}$~cm$^{-2}$.  For the doublets, we give the
total flux in 
both doublet components.}
\tablecaption{Gaussian fits to the line profiles}
\end{deluxetable}}

\clearpage

\begin{table}
\begin{tabular}{ccc}
Phase & $F_{1238}/F_{1242}$ (broad) & $F_{1238}/F_{1242}$ (narrow) \\
\hline
0.057 & $1.10\pm0.20$ &\\
0.092 &  $1.34\pm0.17$ &\\
0.132 &  $1.46\pm0.03$ &\\
0.171 & $1.54\pm0.06$ & $1.44\pm0.35$\\
0.192 & $1.58\pm0.08$ & $1.40\pm0.30$\\
0.211 & $0.94\pm0.11$ & $1.28\pm0.06$\\
0.650      & $1.16\pm0.10$ & $1.23\pm0.23$\\
0.685      & $1.17\pm0.10$ & $1.24\pm0.11$\\
0.725      & $1.11\pm0.09$ & $1.27\pm0.05$\\
0.804      & $1.18\pm0.08$ & $1.30\pm0.08$\\
\end{tabular}
\caption{Doublet ratios for the \ion{N}{5}$\lambda1238.821,1242.804$ lines
from Gaussian fits to the broad and narrow components.}
\end{table}

\setcounter{figure}{0}

\begin{figure}
\caption{\label{fig:plotall}}
\plotone{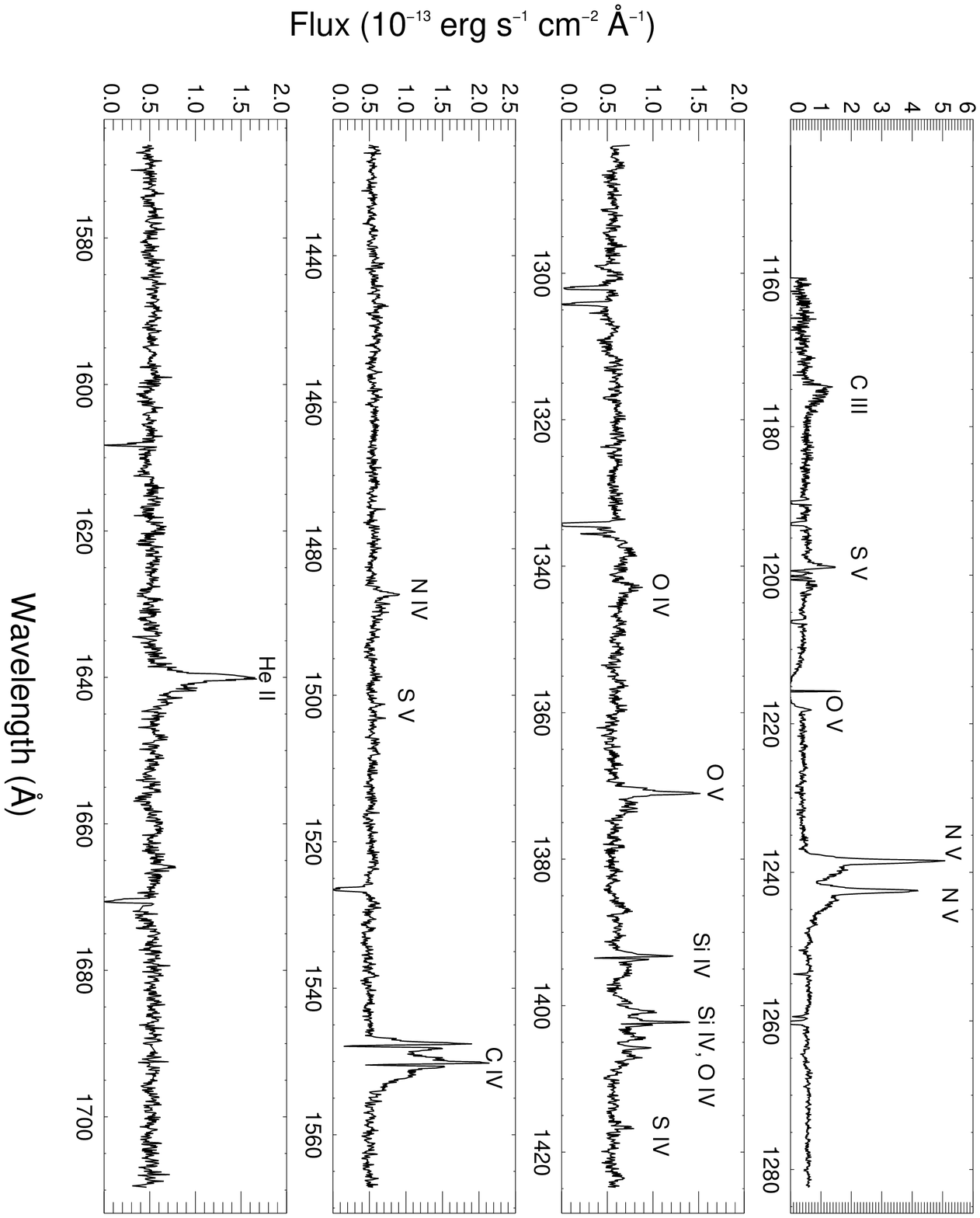}
\end{figure}

\begin{figure}
\caption{\label{fig:gaussn5}}
\plotone{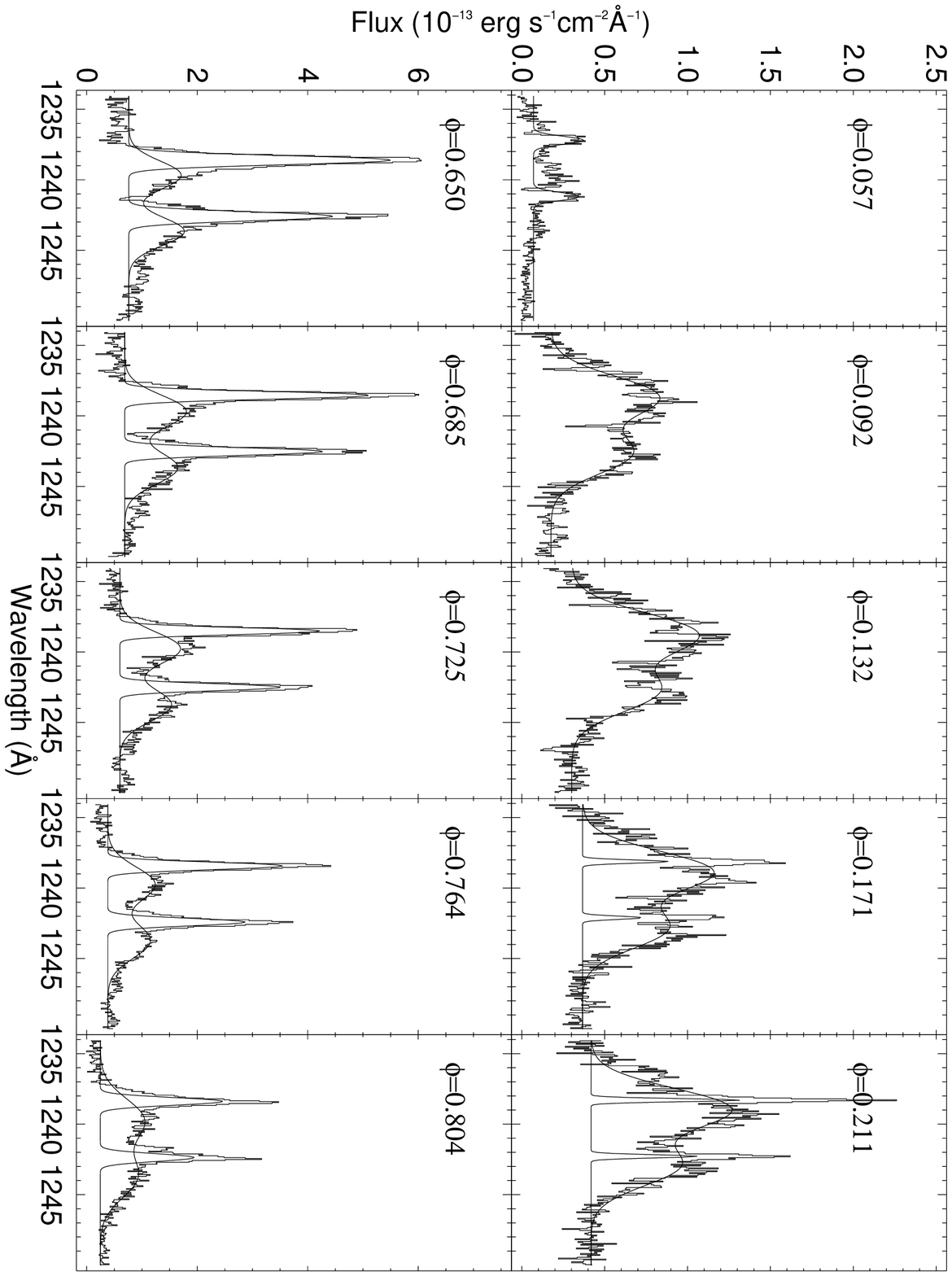}
\end{figure}

\begin{figure}
\caption{\label{fig:gausso5}}
\plotone{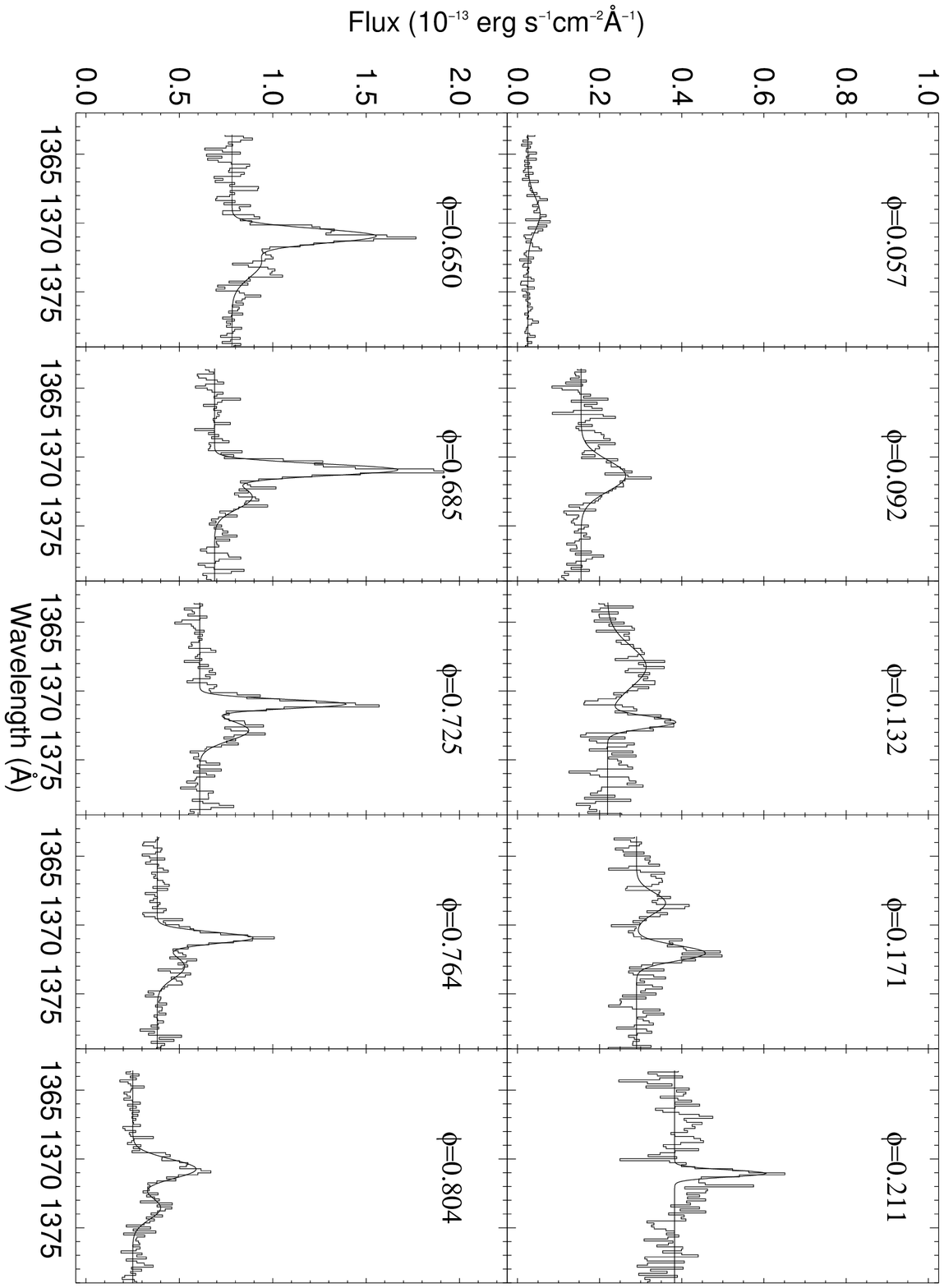}
\end{figure}

\begin{figure}
\caption{}
\plotone{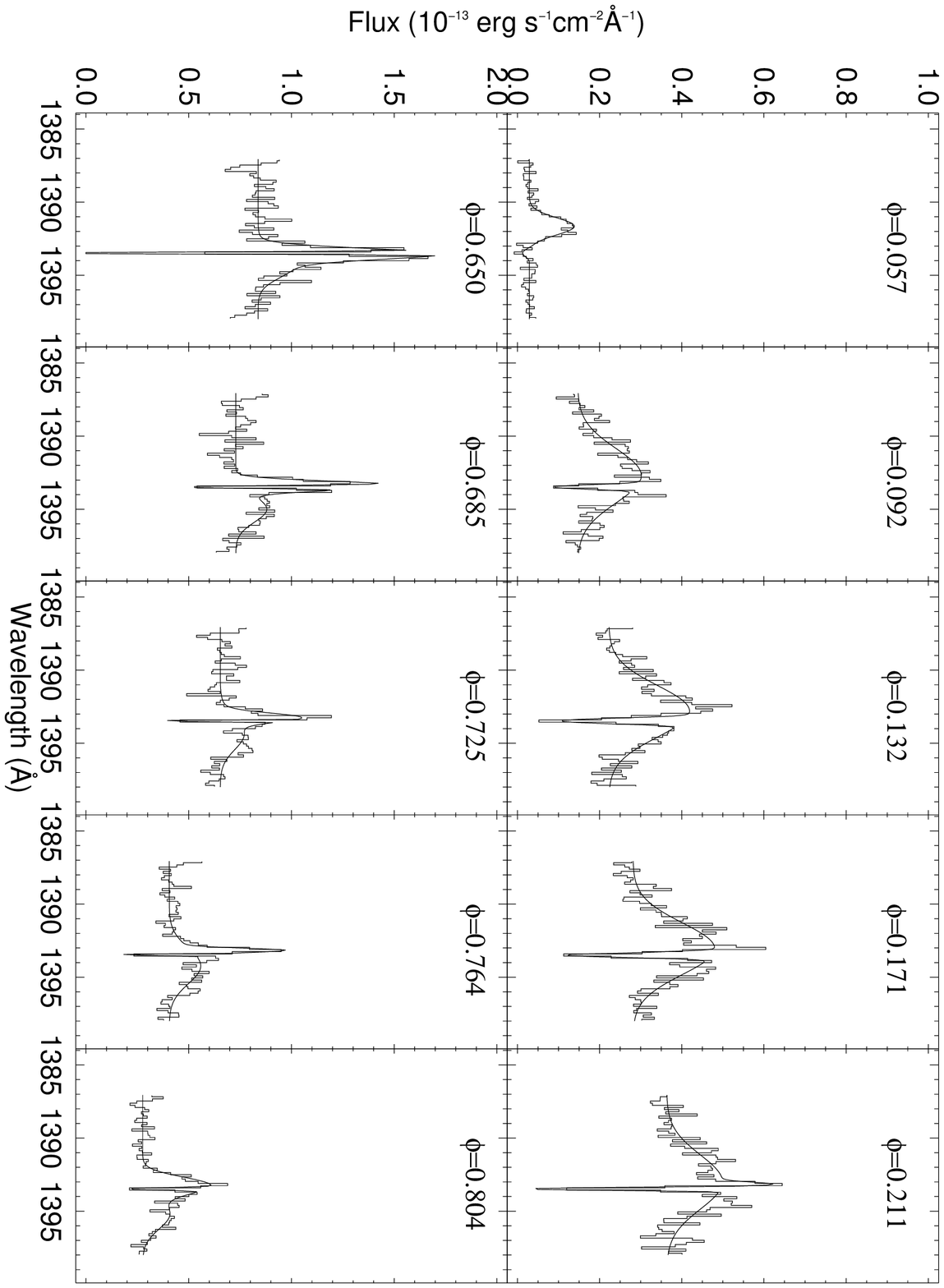}
\end{figure}

\begin{figure}
\caption{}
\plotone{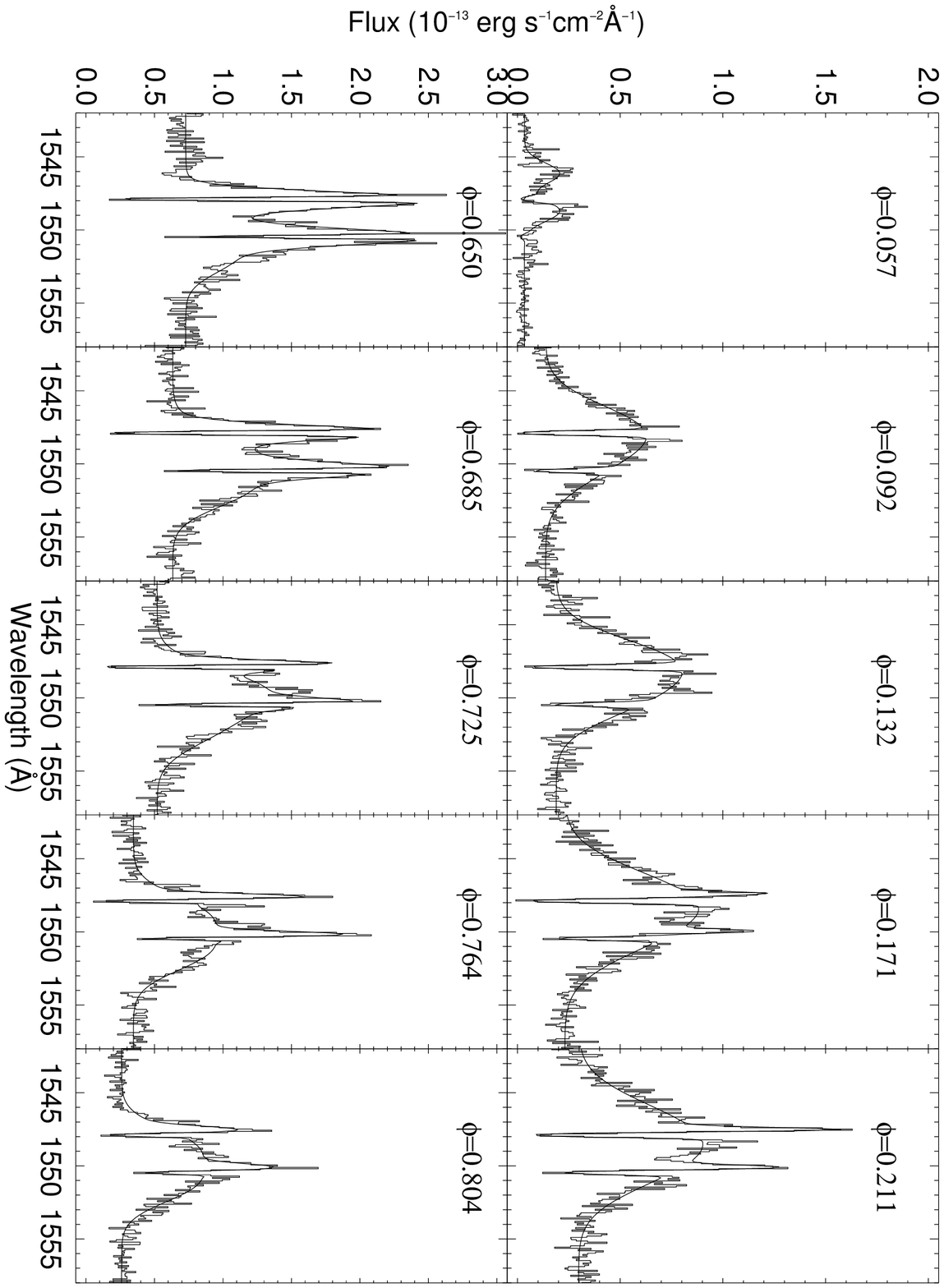}
\end{figure}

\begin{figure}
\caption{\label{fig:gausshe2}}
\plotone{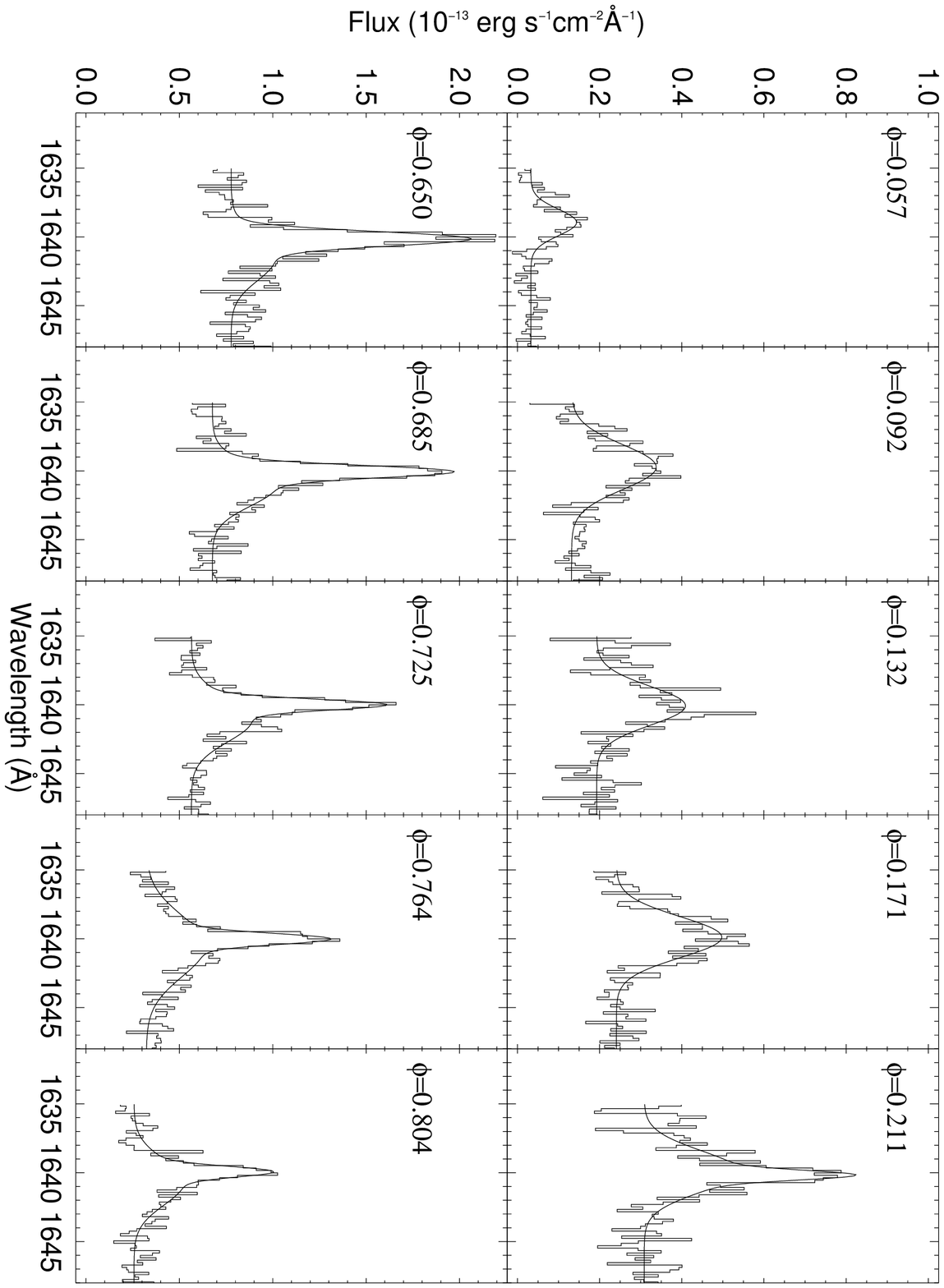}
\end{figure}

\begin{figure}
\caption{\label{fig:linevels}}
\plotone{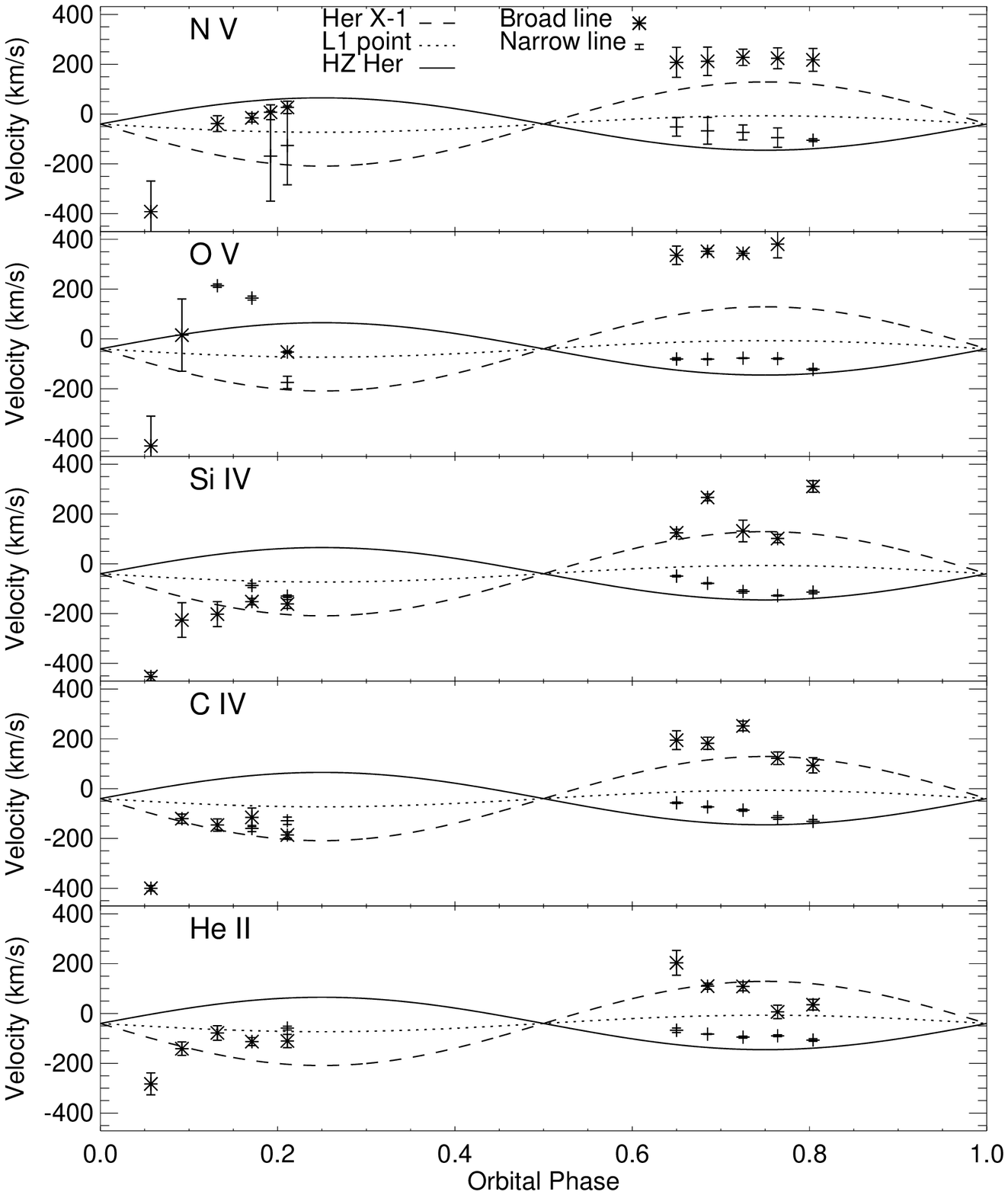}
\end{figure}

\begin{figure}
\caption{\label{fig:fluxn5}}
\plotone{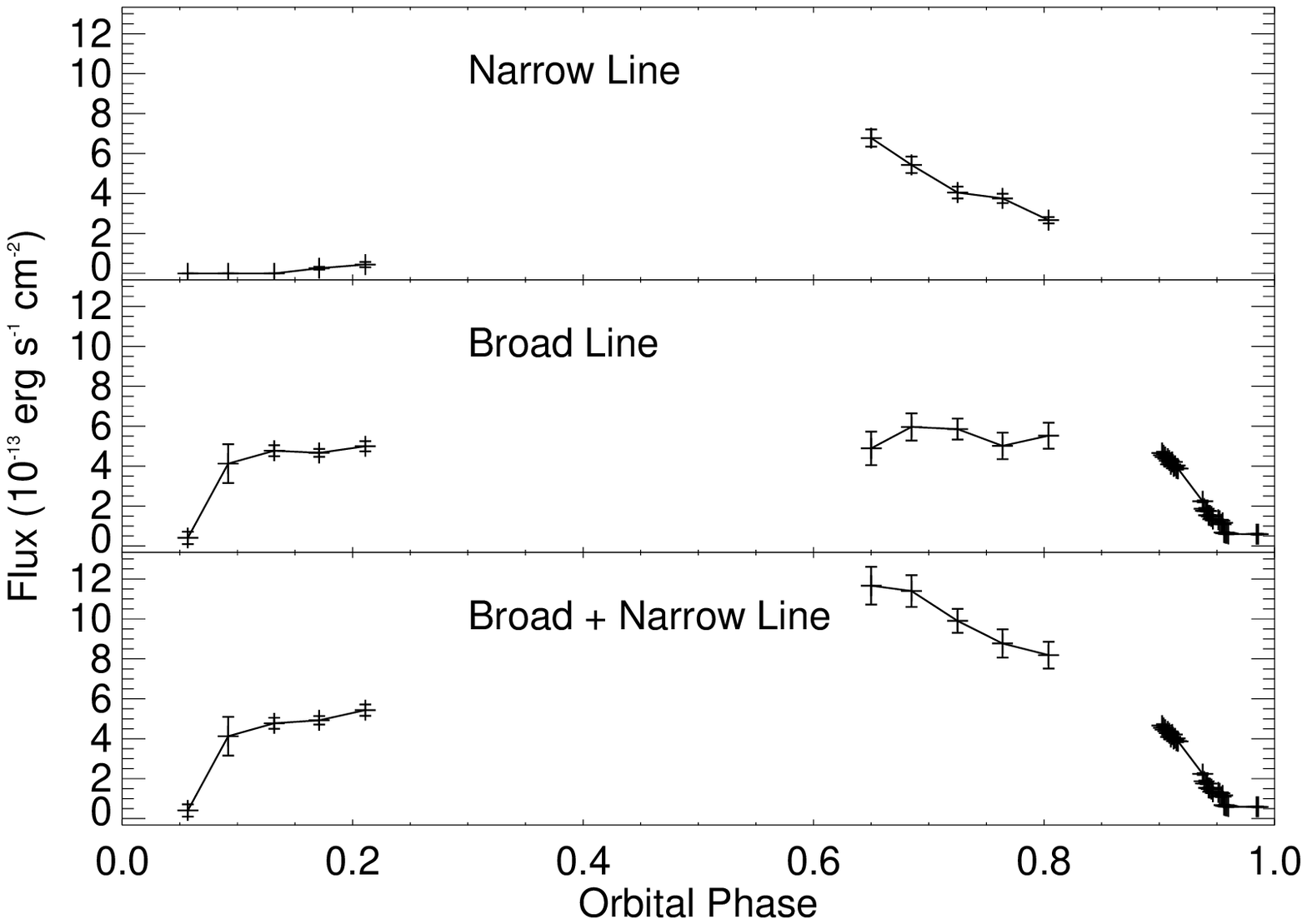}
\end{figure}

\begin{figure}
\caption{\label{fig:iueline}}
\plotone{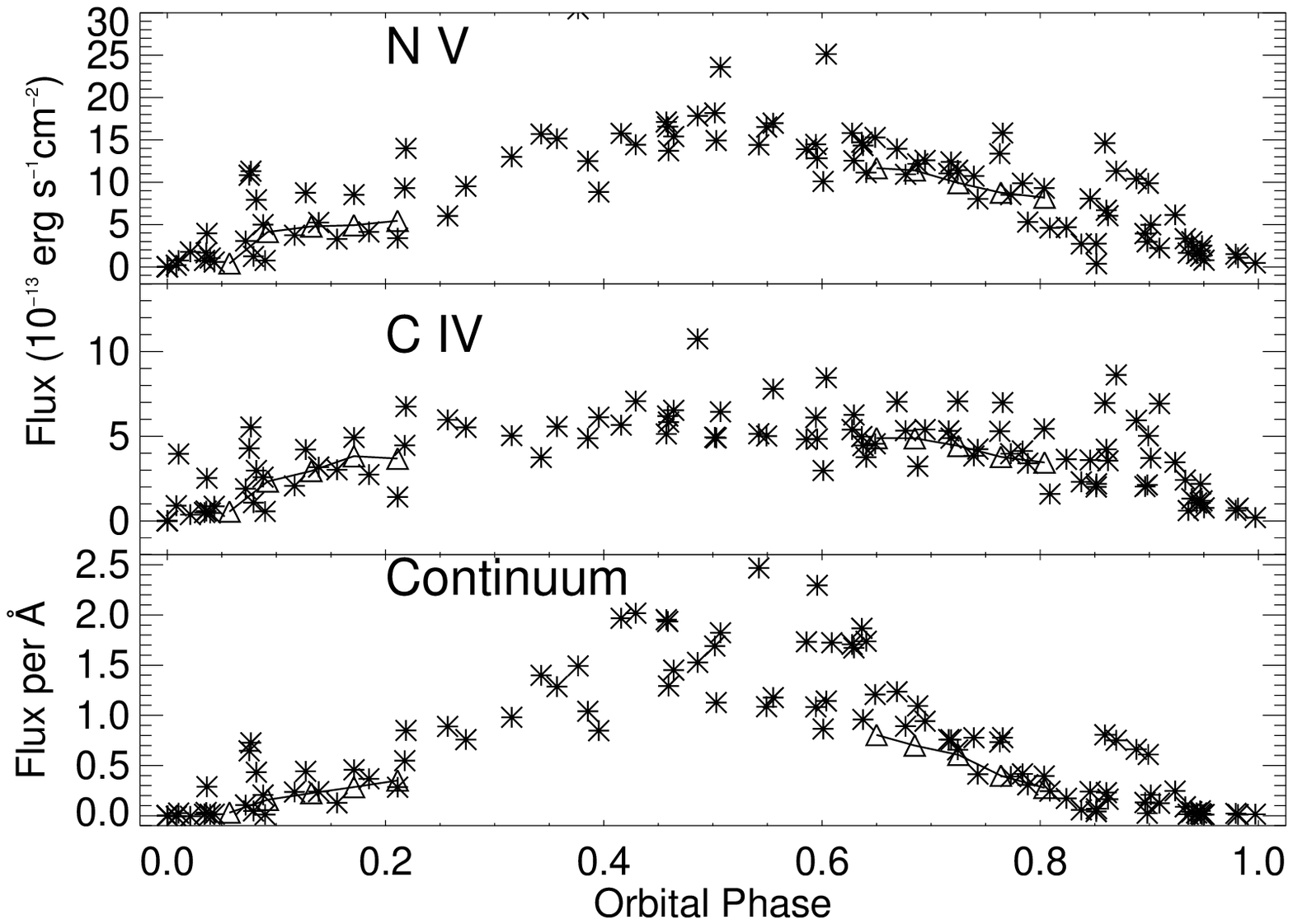}\\
\end{figure}

\begin{figure}
\caption{\label{fig:symmetry}}
\plotone{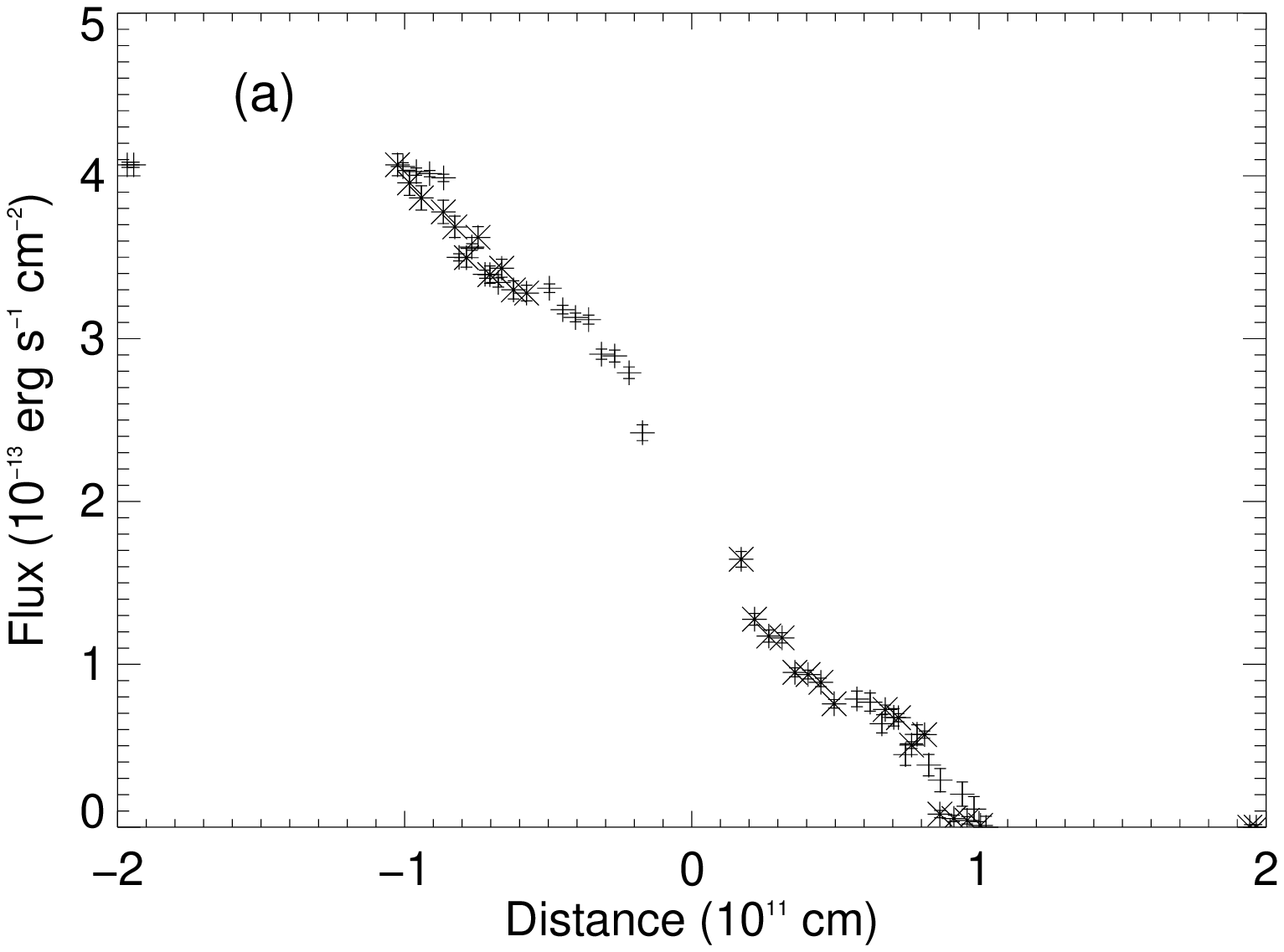}\\
\plotone{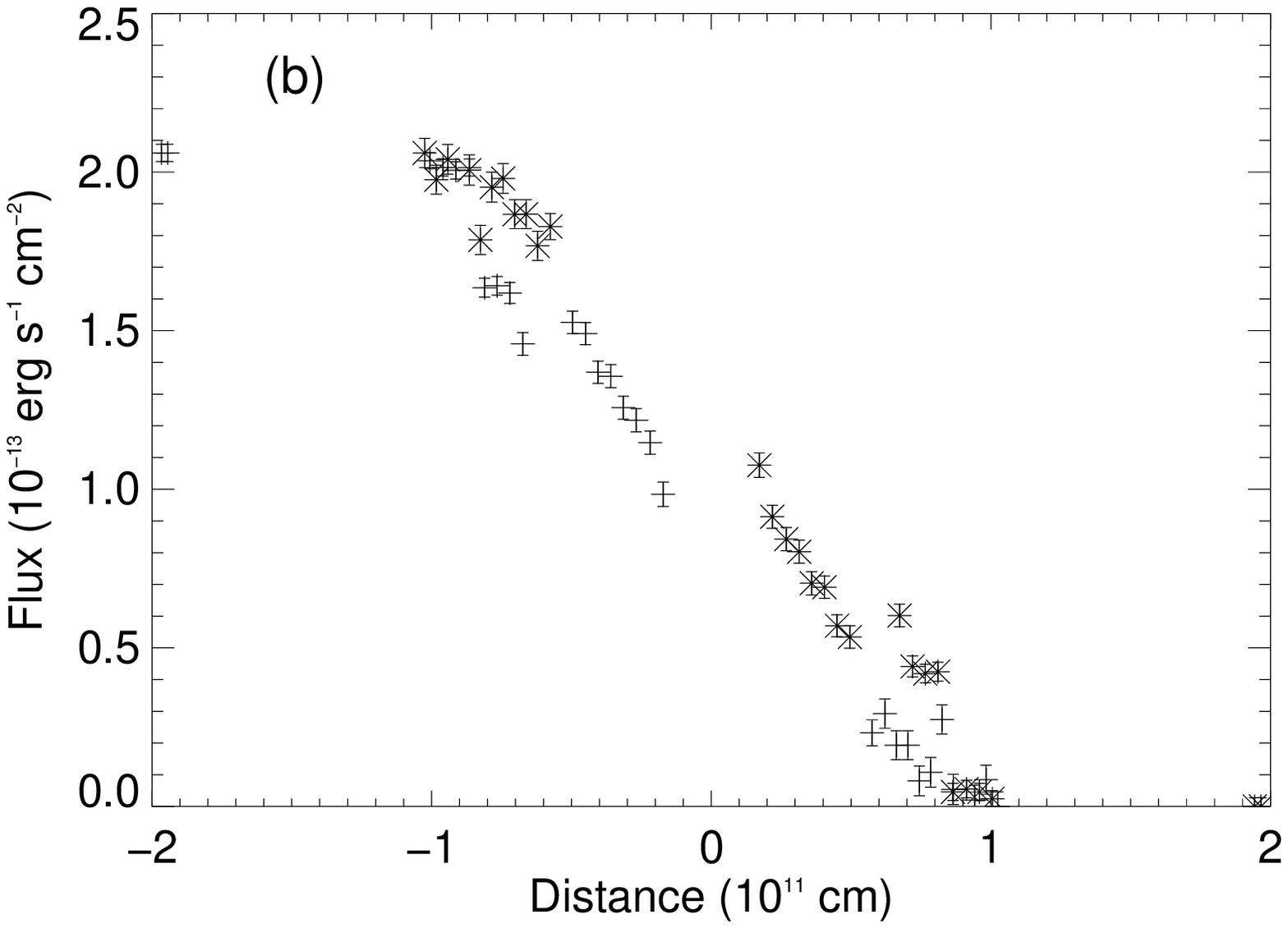}
\end{figure}

\setcounter{figure}{9}

\begin{figure}
\caption{}
\plotone{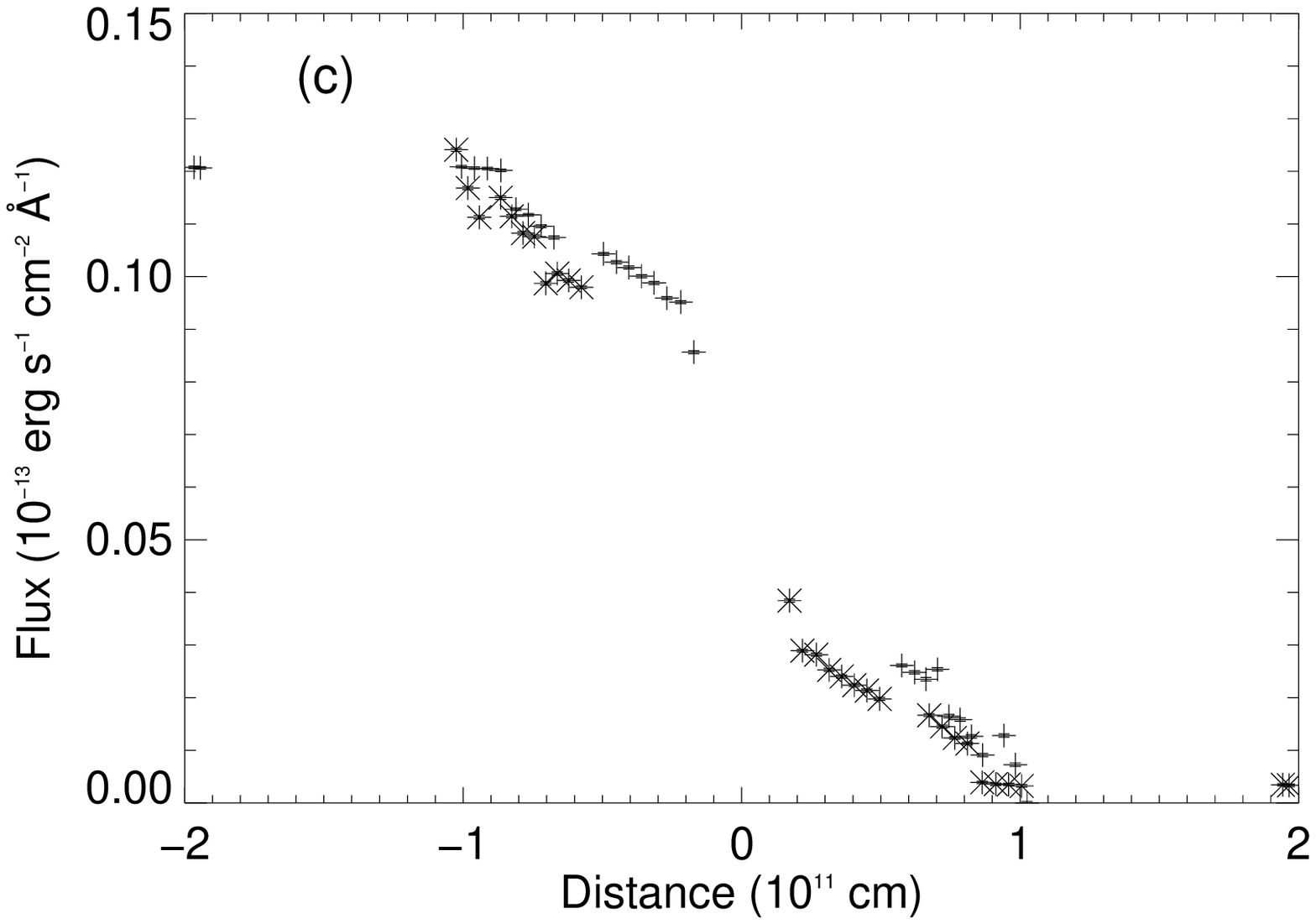}
\end{figure}

\clearpage

\begin{figure}
\caption{\label{fig:lineprof}}
\plotone{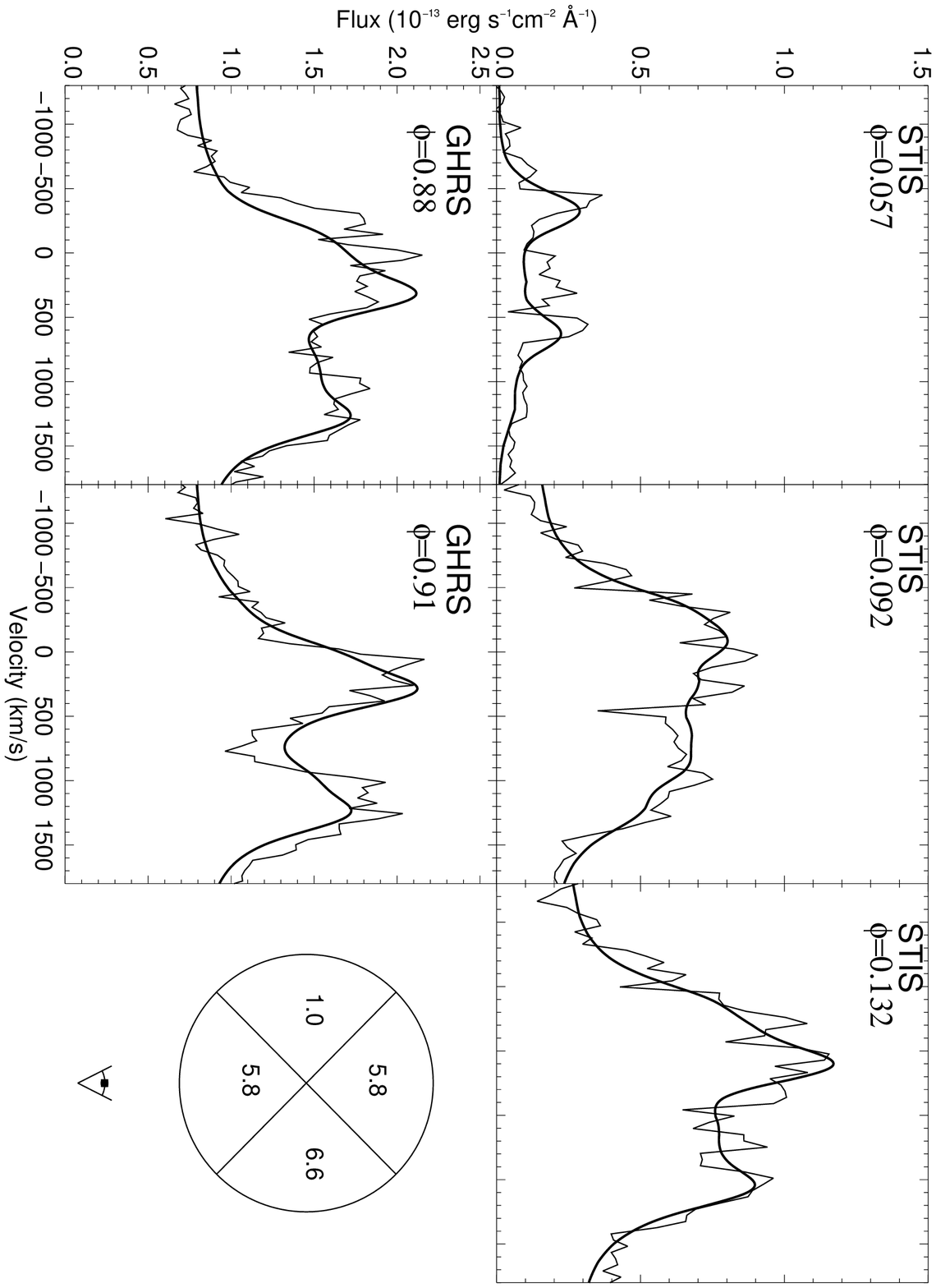}
\end{figure}

\begin{figure}
\caption{\label{fig:o4lines}}
\plotone{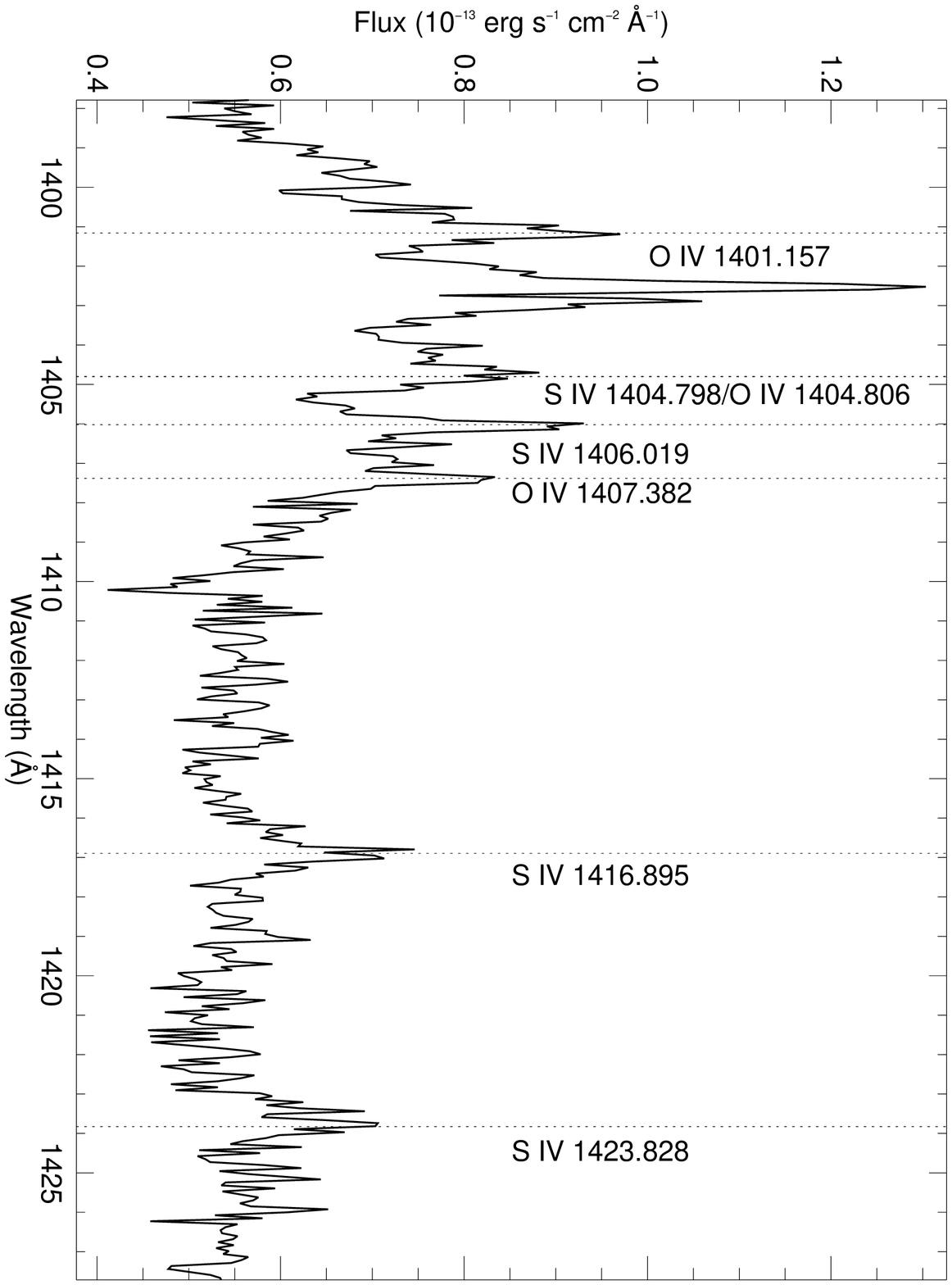}
\end{figure}

\begin{figure}
\caption{\label{fig:manuel}}
\plotone{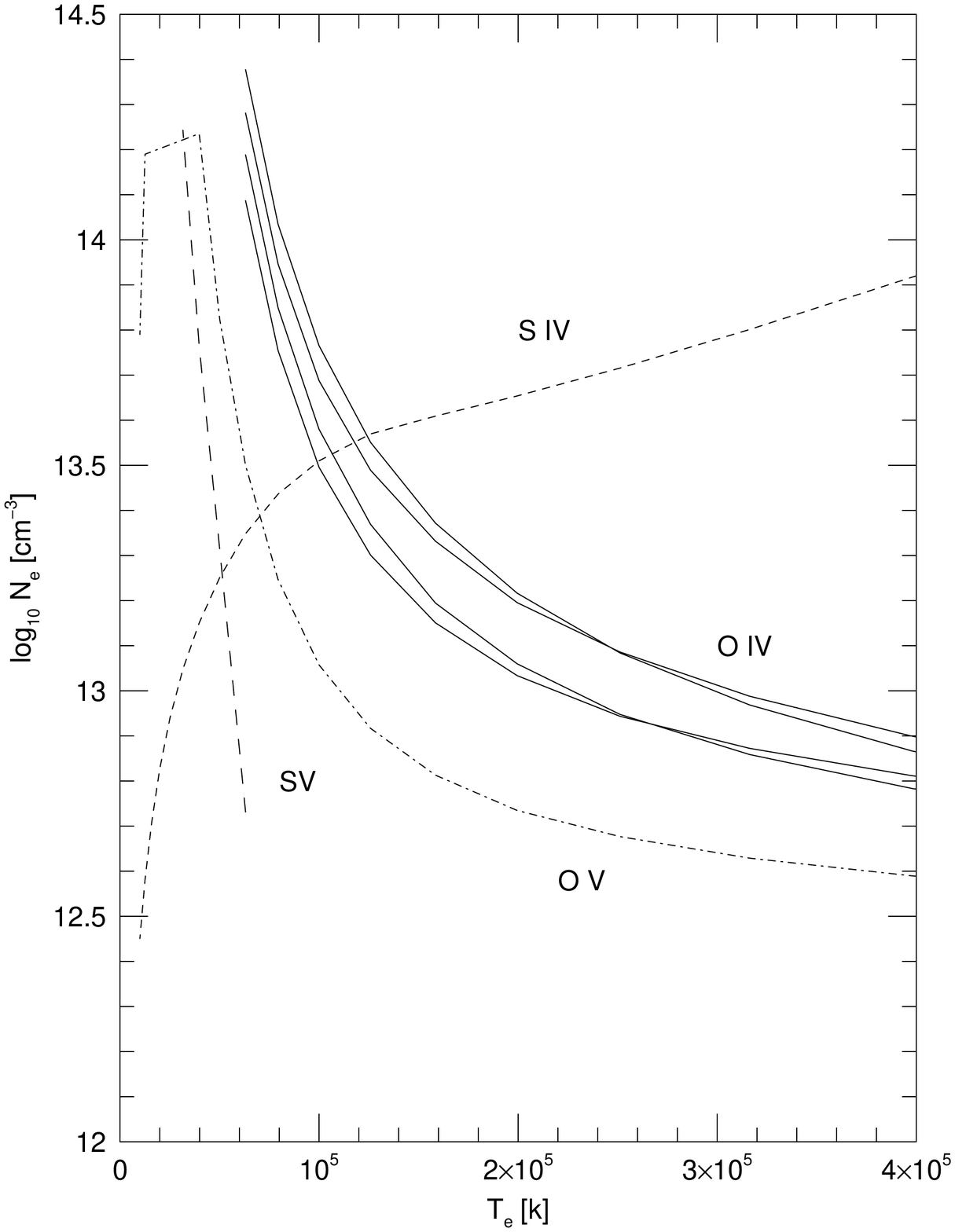}
\end{figure}

\begin{figure}
\caption{\label{fig:abslines}}
\plotone{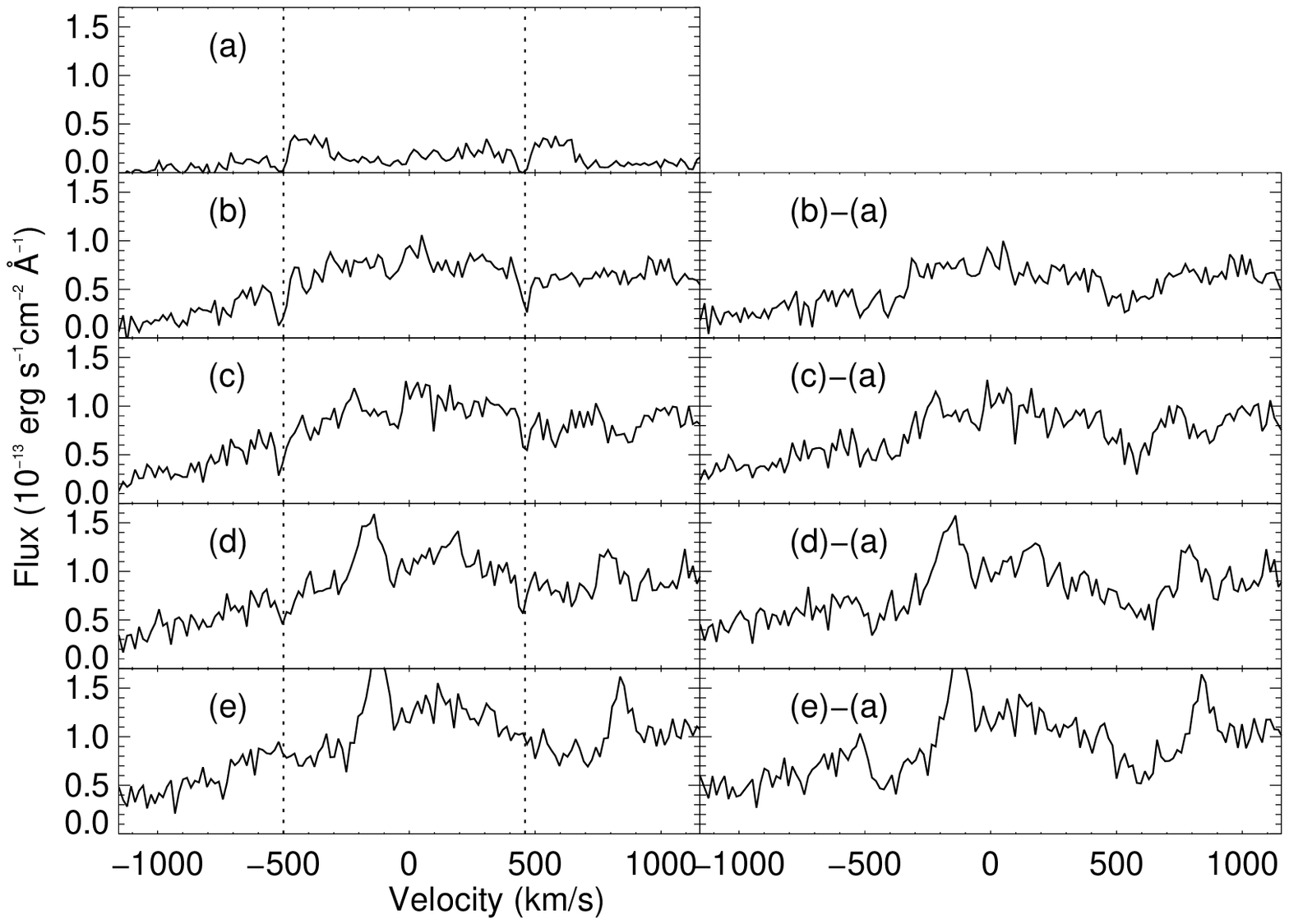}
\end{figure}

\end{document}